\documentclass[preprint,nofootinbib,superscriptaddress,amsmath,amssymb,UTF8]{revtex4-1}
\usepackage{amsmath,mathrsfs,amsbsy,color,graphicx,bm,amsthm,amsfonts}
\usepackage{units}
\usepackage{bbm}
\usepackage{times}
\usepackage{dcolumn}
\usepackage{multirow}
\usepackage{mathrsfs}
\usepackage{amsmath,amssymb,epsfig}
\usepackage{graphicx}
\usepackage{epstopdf}
\usepackage{hyperref}
\DeclareMathSymbol{\shortminus}{\mathbin}{AMSa}{"39}
\usepackage{amssymb}
\usepackage{amsbsy}
\usepackage{graphicx}
\usepackage{amsmath}
\usepackage{times,txfonts}
\newcommand{\pp}{\partial}
\begin{document}

\title{Quasinormal Modes and Dynamical Evolution of Scalar Fields in the Einstein–Bumblebee Theory with a Cosmological Constant}

\author{Hao Hu}
\email[]{huhao91@csu.edu.cn} 
\affiliation{ Department of Oncology, The Third Xiangya Hospital, Central South University, Changsha, 410013, Hunan, China}

\author{Guoxiong Zhu}
\email[]{846723141@qq.com (Corresponding authors)} 
\affiliation{Radiotherapy Department, Hunan Hospital of Integrated Traditional Chinese and Western Medicine, Changsha, 410006, Hunan, China}
\affiliation{Institute of Nuclear Science and Technology , Sichuan University, Chengdu, 610064, Sichuan, China}

\vspace*{0.2cm}
\begin{abstract}
In this paper, we investigate the quasi normal modes (QNMs) of static spherically symmetric black holes in the Einstein—Bumblebee gravity model, taking into account the effects of the field mass and the cosmological constant.
Through separation of the angular components, the scalar field perturbations outside the black hole are reduced to a purely radial main equation. 
We calculated the quasinormal modes using the matrix method and the WKB approximation, and also studied the dynamical evolution of the purely radial main equation using the finite difference method in the time domain.
The eigenfrequencies of the waveforms from the time-domain evolution are fitted to cross-validate the frequency-domain results.
The Lorentz-violating parameter $ \ell $, cosmological constant $ \Lambda $, and scalar field mass $ \mu M $ affect QNMs.
Specifically, increasing $ \ell $ decreases both real and imaginary parts of the monopole modes, but in the dipole and quadrupole modes, the real part remains nearly unchanged while the imaginary part drops rapidly; rising $ \Lambda $ reduces both parts of QNMs; increasing $ \mu M $ raises the real part and lowers the imaginary part. 
Time-domain analysis confirms these findings, clarifying how Lorentz symmetry breaking impacts QNMs in ds Sitter spacetime.
\\


\end{abstract}

\vspace*{0.5cm}

\maketitle
 
\section{Introduction}

In the context of general relativity, Lorentz invariance is a fundamental principle that underpins both general relativity and the standard model of particle physics \cite{Magueijo2002,Baojiu2011,Heisenberg2019,Martinez2020}. 
However, at high energy scales, particularly within quantum gravity, Lorentz invariance may no longer hold \cite{Ashtekar2021,Campiglia2017}. 
This limitation has prompted a growing interest in investigating Lorentz violation as a key path toward deepening our understanding of the fundamental principles of modern physics. 
The standard model extension is an effective field theory that integrates general relativity with the standard model, incorporating dynamic curvature modes and terms that account for Lorentz-violating effects at the Planck scale \cite{Colladay1998}.
Under the influence of the local Lorentz force, violation of Lorentz symmetry can be detected by nonzero vacuum expectation values of these indices \cite{Roberts2017,Yang:2023wtu,Duan:2023gng,Filho:2023ycx,Liu:2024oas,Liu:2024lve,AraujoFilho:2024rcr,Hosseinifar:2024wwe,Chen:2024ilt}.
A notable approach to exploring Lorentz violation is the Bumblebee model, a variant of Einstein's aether theory, characterized by a potential function that minimally rolls towards its vacuum expectation values \cite{Berglund2013}. 
Initially proposed by Kostelecky and Samuel in 1989 as a model for spontaneous Lorentz-symmetry breaking \cite{Kostelecky1989}, this model has since been studied for its implications in various contexts. 
Casana et al. identified an exact solution in the Bumblebee gravity model that resembles the Schwarzschild solution, examining its effects on general relativistic phenomena such as planetary perihelion progression, light bending, and time delays caused by curvature \cite{Casana2018}. 
Additionally, the solution has been extended to incorporate a cosmological constant by Maluf \cite{Maluf2021}, a significant advancement that enables a deeper exploration of the effects of Lorentz violation in both de Sitter and anti-de Sitter spacetimes. 
Given the accelerated expansion of the universe, current models suggest that the universe is asymptotically de Sitter. 
Thus, this paper aims to explore the effects of Lorentz violation on the dynamical properties of black holes in an asymptotically de Sitter spacetime.
Furthermore, within this framework, more black hole solutions have been solved \cite{Filho:2022yrk,Ovgün2019,Ding2020,Ding2023,Ding:2021iwv}. 
Since then, this solution has served as a foundation for further research into related topics, including quasinormal modes (QNMs) \cite{Oliveira2021,Liu2023,Malik:2023bxc,Mai:2024lgk}, black hole shadows \cite{Jha:2020pvk,Kuang2022,Gao2024,Pantig:2024ixc,Liu:2024axg,Liu:2024iec,Liu:2024soc,Afrin:2024khy}, photon trajectories \cite{Lambiase2023}, and studies on entanglement degradation \cite{Liu:2024wpa,Liu:2025bpp} or particle creation \cite{AraujoFilho:2025hkm}, as well as the metric affine case \cite{Filho:2023etf,Amarilo:2023wpn,Lambiase:2023zeo,AraujoFilho:2024ykw,Gao:2024ejs,Heidari:2024bvd,Jha:2024fnh}, among others \cite{AraujoFilho:2024iox,Filho:2024isd,Liu:2024oeq,Ji:2024aeg}.

On the other hand, in black hole physics, interactions between black holes and matter fields are crucial for understanding black hole stability, testing the no-hair theorem \cite{Bhattacharya2007}, and developing new black hole solutions \cite{Herdeiro2016}. 
To study these interactions, researchers often use the linear approximation of black hole perturbation theory \cite{Calcagni2017,Mukohyama2022}. Black hole perturbations can be divided into two types. 
The first type involves perturbations of the black hole's metric itself, where gravitational radiation from black hole disturbances exceeds that caused by external fields \cite{Del2021}, making it a key measure of black hole stability. 
The second type involves perturbations within the ambient spacetime of the black hole, arising from disturbances caused by matter fields. 
These perturbations reflect the dynamic behavior of matter fields near the black hole and offer insights into the interactions between black holes and their environment. 
Similar to ripples on water caused by external forces, spacetime oscillations propagate as gravitational waves, with energy gradually dissipating during transmission. Perturbative fields in black hole spacetimes typically undergo three stages: initial bursts, QNMs, and power-law tails \cite{Destounis2020}. 
The latter two stages, linked to the intrinsic properties of the black hole, are of particular interest. 
These stages, often described as the "ringdown" of black holes, exhibit characteristics akin to the acoustic signatures of black holes \cite{Chakrabarti2001}. 
By analyzing these signatures, researchers can identify and categorize different types of black holes.

This paper investigates the QNMs and the dynamic waveform evolution in de Sitter black hole spacetimes within the Einstein—Bumblebee gravitational model and contrasts these outcomes with the results from traditional Schwarzschild-de Sitter spacetimes \cite{Zhidenko2004}. 
The impact of different Lorentz violation parameters on the properties of black hole spacetimes is explored. 
Sec. \ref{sec2} of this manuscript provides a brief overview of the Einstein—Bumblebee gravity model and its underlying geometrical background and derives equations of motion for the scalar field. 
Sec. \ref{sec3} describes the WKB method \cite{Konoplya2003}, employs numerical techniques to find the scalar field QNMs, and presents the associated numerical results. 
Sec. \ref{sec3} briefly introduces the matrix method \cite{Lin2017} and the WKB method \cite{Konoplya2003} to solve the quasinormal modes of the scalar field and presents the corresponding numerical calculation results.
In Sec. \ref{sec4}, we delve into the dynamical evolution of the scalar field under a small cosmological constant, revealing significant late-time behavior characterized by power-law tails. 
The final section offers a perspective on potential future research directions for this topic. 

\section{Einstein—Bumblebee Gravity Model}\label{sec2}
The action describing Lorentz symmetry breaking is expressed as:
\begin{equation}
S = \int d^4x \sqrt{-g} \left[ \frac{1}{2\kappa} \left( R - 2\Lambda \right) + \xi B^\mu B^\nu R_{\mu\nu} - V(B^\mu B_\mu \pm b^2) + \mathcal{L}_M \right], \label{eq:action}
\end{equation}
where $\kappa$ is the gravitational coupling constant, $R$ is the Ricci scalar, $\Lambda$ is the cosmological constant, and $\xi$ is a real coupling constant determining the nonminimal gravitational interaction of the Bumblebee field $B_\mu$. 
The field strength of the Bumblebee field is defined as $B_{\mu\nu} = \partial_\mu B_\nu - \partial_\nu B_\mu$, and $\mathcal{L}_M$ describes the matter. 
The potential $V$ is given by $ V=V(B_\mu B^\mu\pm b^2) $, which provides a nonzero vacuum expectation value $ \langle B_\mu \rangle $. 
Importantly, the selected potential $V$ is designed to ensure that the Bumblebee field $\langle B_\mu \rangle=b_\mu$ has a nonzero vacuum expectation value, which triggers the spontaneous breaking of Lorentz symmetry. 
The potential $V$ reaches a minimum at $B^\mu B_\mu \pm b^2 $, where $b$ is a positive real constant. 
The $\pm$ sign indicates that, depending on whether the Bumblebee field is timelike or spacelike, different scenarios may arise.

By varying the action with respect to the metric tensor $g_{\mu\nu}$ and the Bumblebee field $B^\mu$ separately, the equations of motion for the vacuum gravitational field and the Bumblebee field can be derived:
\begin{equation} \label{eq:field_eq}
R_{\mu\nu}-\frac{1}{2}g_{\mu\nu}(R-2\Lambda)=\kappa T^M_{\mu}+\kappa T^B_{\mu\nu},
\end{equation}
\begin{equation}\label{eq:bumblebee_eq}
\nabla^\mu B_{\mu\nu}=2V'B_\nu-\frac{\xi}{\kappa}B^\mu R_{\mu\nu}.
\end{equation}
Here, $T_{\mu\nu}^M$ is the energy—momentum tensor of the matter field and
\begin{equation}
T^B_{\mu\nu}=-B_{\mu\nu}B^{\alpha}_{~\nu}-\frac{1}{4}B_{\alpha \beta} B^{\alpha\beta}+2V'B_\mu B_\nu
-Vg_{\mu\nu}+\frac{\xi}{\kappa}\mathcal{B}_{\mu\nu},
\end{equation}
is the Bumblebee energy—momentum tensor with 
\begin{equation}
\begin{aligned}
\mathcal{B}_{\mu\nu}=&\frac{1}{2}B^{\alpha}B^\beta R_{\alpha\beta}g_{\mu\nu}-B_{\mu}B^{\alpha}R_{\nu\alpha}-B_{\nu}B^{\alpha}R_{\mu\alpha}
+\frac{1}{2}\nabla_\alpha\nabla_\mu B^\alpha B_\nu\\
&+\frac{1}{2}\nabla_\alpha\nabla_\nu B^\alpha B_\mu-\frac{1}{2}\nabla^2 B_\mu B_\nu-\frac{1}{2}g_{\mu\nu}\nabla_\alpha\nabla_\beta B^\alpha B^\beta.
\end{aligned}
\end{equation}

By solving this system of differential equations, one can obtain curved spacetime metrics with asymptotically de Sitter (dS) or anti-de Sitter (AdS) properties under the influence of the Bumblebee field. 
These can also be referred to as Einstein—Bumblebee black hole metrics, and their line element is given by:
\begin{equation} \label{eq:metric}
ds^2 = -F(r) dt^2 + (1+\ell)F(r)^{-1} dr^2 + r^2 d\theta^2+r^2\sin^2\theta d\varphi^2,
\end{equation}
where $F(r)$ represents the metric function, specifically given by $ F(r)=1-\frac{2M}{r}-\frac{(1+\ell)\Lambda_{eff}}{3}r^2 $.
The factor denoting Lorentz symmetry breaking, symbolized by $\ell$, indicates the level of spacetime symmetry disruption.
This factor is determined jointly by the coupling constant $ \xi $ and the nonzero vacuum expectation value of the Bumblebee field $ b $, with the formula $ \ell = \xi b^2 $. 
Notably, $\Lambda_{eff}$ is the effective cosmological constant, which is defined as $\Lambda_\text{eff} = \Lambda - \xi b^2$, and $\lambda$ is the Lagrange multiplier field \cite{Oliveira2021}. 
Together with the constraint $B^\mu B_\mu \pm b^2 = 0$, it ensures that the cosmological constant can be reasonably accommodated within the Bumblebee gravity model.
In other words, a direct relationship exists between the effective cosmological constant and the cosmological constant, as defined by the constraint:
\begin{equation}\label{eq:cosmo_const}
\Lambda=(1+\ell)\Lambda_\text{eff}.
\end{equation}
The metric function in references can thus be rewritten as $F(r)=1-\frac{2M}{r}-\frac{\Lambda r^2}{3}$, which is consistent with the Schwarzschild—de Sitter case. 
However, the asymptotic behavior of the metric at infinity is not consistent with the Schwarzschild—de Sitter case, and the influence of Lorentz violation also needs to be considered. 
When the cosmological constant $\Lambda$ is zero, the resulting metric coincides with the results obtained by Casana in 2018 \cite{Casana2018}. 
When the Lorentz violation parameter $\ell = 0$, the metric simplifies to Schwarzschild—(A)dS. 
The incorporation of the cosmological constant modifies the structure of spacetime, resulting in a dual horizon configuration that encompasses both the black hole event horizons and cosmological horizons. 
This means that, the metric function can be reformulated as \cite{Jing:2003wq,Liu:2023uft,Liu:2024lbi}:
\begin{equation}\label{eq:horizons}
F(r)=\frac{\Lambda}{3}\left(1-\frac{r_h}{r}\right)\left(r_c-r\right)\left( r+r_h+r_c\right).
\end{equation}
In this manuscript, $r_\text{h}$ and $r_\text{c}$ represent the black hole event horizon and the cosmological horizon, respectively\cite{Zhidenko2004,Yu2022}. 
Our universe is situated between these two horizons, specifically in the exterior communication region where $r_h < r < r_c$. 
This region is causally disconnected from both the interior region of the black hole, where $r < r_h$, and the external region of the universe, where $r > r_c$. 
Consequently, the exterior communication region is causally independent in its structure.

\subsection{Massive Scalar Field Perturbation Equation}
\begin{figure}[h]
\centering
\includegraphics[width=0.8\textwidth]{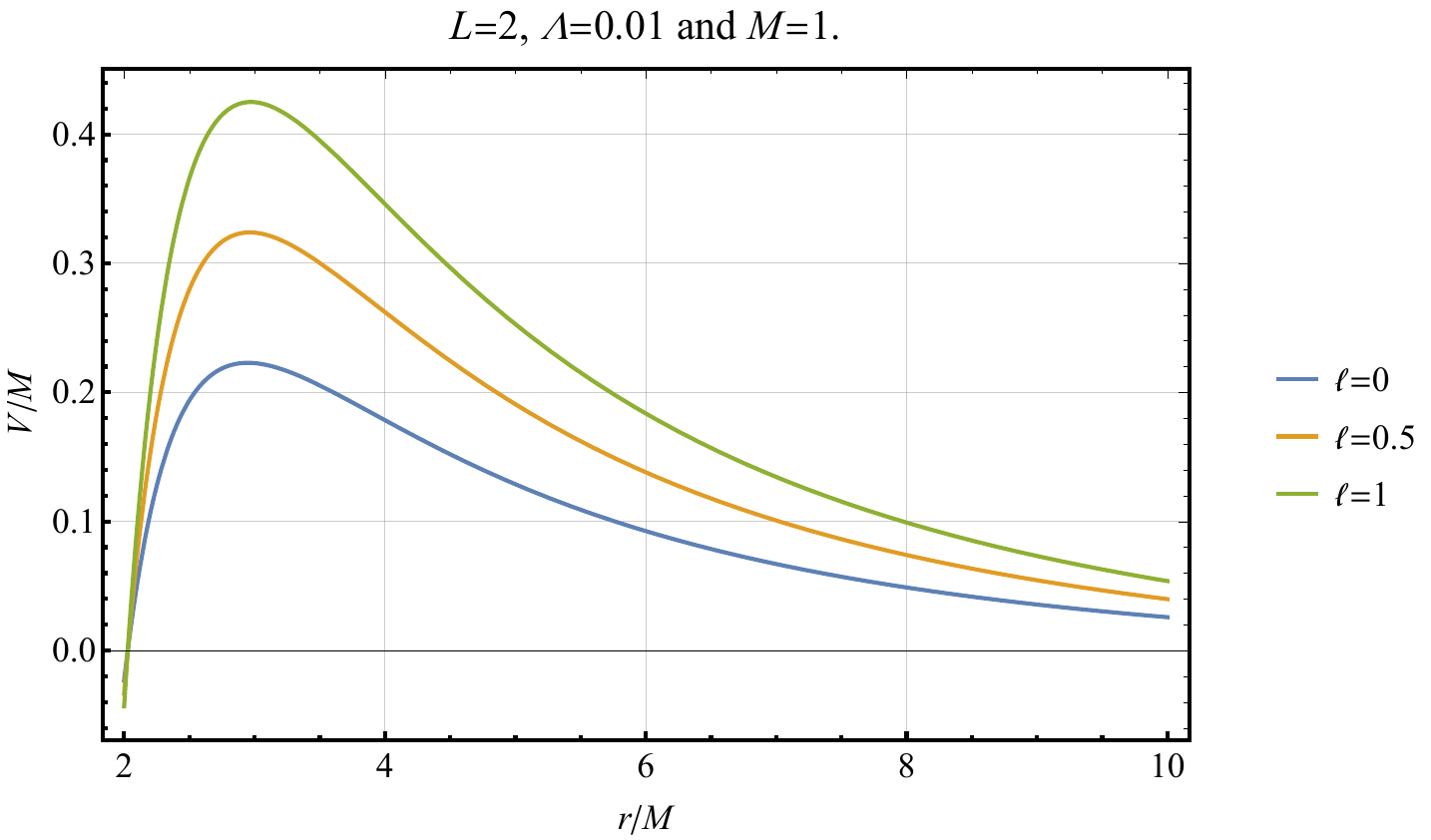} 
\caption{The effective potentials for \( L=2 \) and \( \Lambda=0.01 \) are shown.}
\label{fig11}
\end{figure}

The equation of motion for a massive scalar field is described by the Klein—Gordon equation:
\begin{equation}
\frac{1}{\sqrt{-g}} \partial_\mu \left( \sqrt{-g} g^{\mu\nu} \partial_\nu \Phi \right) = \mu^2\Phi, \label{eq:klein_gordon}
\end{equation}
where $g$ denotes the determinant of the metric. 
We employ spherical harmonics $Y^{Lm}(\theta, \phi)$ to separate variables in the scalar function $\Phi$:
\begin{equation}\label{eq:separation}
\Phi(t, r, \theta, \phi) = \frac{1}{r} \psi(t,r) Y^{Lm}(\theta, \phi). 
\end{equation}
Inserting equation \eqref{eq:separation} into equation \eqref{eq:klein_gordon} yields a mixed equation for the radial and angular components. 
By utilizing the associated Legendre equation in spherically symmetric coordinates, we can rewrite the angular part as an eigenvalue expression as follows:
\begin{equation}\label{eq:angular_part}
\left[\frac{1}{\sin\theta} \frac{\partial}{\partial\theta} \left( \sin\theta \frac{\partial }{\partial\theta} \right) + \frac{1}{\sin^2\theta} \frac{\partial^2 }{\partial\phi^2}\right] Y^{Lm} = L(L+1) Y^{Lm}, 
\end{equation}
where $L$ represents the angular quantum number \cite{Zhidenko2004}. 
The angular component from the mixed equation can be eliminated by substituting into the eigenvalue equation \eqref{eq:angular_part}, resulting in a purely radial master equation given by:
\begin{equation}\label{eq:radial_master}
\left[ F(r)^2 \frac{\pp^2}{\pp r^2}-(1+\ell)\frac{\pp^2}{\pp t^2}+F(r)F'(r)\frac{\pp}{\pp r}-V(r) \right]\psi(t,r)=0,
\end{equation}
where the prime notation ($'$) indicates differentiation with respect to the radial coordinate, and $V(r)$ represents the effective potential for scalar perturbations, expressed as
\begin{equation}
\begin{aligned}
V(r)=&\frac{F(r)}{r^2}\left[(1+\ell)L(L+1)+(1+\ell)r^2\mu^2+\frac{2M}{r}-\frac{2\Lambda r^2}{3}  \right].
\end{aligned}
\end{equation}
The equation can be transformed into a form akin to the Schrödinger equation by applying a tortoise coordinate transformation. 
Continuing from the previous transformation,
\begin{equation}
\begin{aligned}
r_*&=\int \frac{1}{F(r)} dr=\eta_h\ln \left(r-r_h\right)+\eta_c\ln\left(r_c-r\right)+\eta_i\ln\left(r+r_h+r_c\right),
\end{aligned}
\end{equation} 
where \( \eta_i = -\eta_h - \eta_c \), and  
\begin{equation}
\begin{aligned}
&\eta_h=\frac{3r_h}{\Lambda}\left(r_c-r_h\right)^{-1}\left(r_h+2r_c\right)^{-1},\\
&\eta_c=\frac{3r_c}{\Lambda}\left(r_h^2+r_h r_c-2r^2_c \right)^{-1}.
\end{aligned}
\end{equation}
Substituting the transformed time coordinate \( t_* = r/\sqrt{1 + \ell} \), the reformulated master equation is expressed as:  
\begin{equation}\label{eq:schrodinger}
\left[\frac{\pp^2}{\pp r^2_*}-\frac{\pp^2}{\pp t^2_*}-V(r) \right]\psi(t,r)=0.
\end{equation}
We can use standard methods to solve the eigenfrequencies of the equation, such as the matrix method \cite{Lin20171,Lin2017,Lin2019}.
It is worth noting that if the field mass is not considered, we can also use the WKB method to verify the results \cite{Iyer1987}.
Meanwhile, we plotted the potential \( V(r) \) against \( r \) for various parameters in Fig. \ref{fig11}. 
From the potential plot, when $ \mu=0 $, we can see that it satisfies the criteria required by the WKB method.

In such black hole spacetime systems, the dissipation of energy primarily arises from the imposition of fully absorbing boundary conditions. 
Specifically, this includes two requirements: at the black hole horizon, the perturbation must be an incoming wave, and at infinity, it should be an outgoing wave. 
These boundary conditions apply to asymptotically flat spacetimes. 
More precisely, the boundary conditions encompass two stipulations: perturbations must manifest as ingoing waves at the black hole horizon and as outgoing waves at infinity.
However, in de Sitter or anti-de Sitter spacetimes, the choice of boundary conditions differs. 
In de Sitter spacetimes, the de Sitter horizon replaces the role traditionally played by spatial infinity; in anti-de Sitter spacetimes, the requirement for the convergence of solutions at spatial infinity supersedes the condition for outgoing waves.
These distinct boundary conditions reflect the unique properties of different types of spacetimes. 
If equations in the frequency domain are considered, one can separate variables in the wave function, allowing for $ \psi(t, r) = \psi(r) e^{-i\omega t} $, the socalled harmonic ansatz, leading to
\begin{equation}
\frac{d^2}{dr_*^2}\psi(r)+\left[(1+\ell)\omega^2-V(r) \right]\psi(r)=0.
\end{equation}
The quasinormal mode problem fundamentally represents a frequency eigenvalue problem for the wave equation. 
The real and imaginary parts of the quasinormal mode frequency $\omega=\omega_R+i\omega_I$ each have distinct physical meanings expressed by $ e^{-i\omega t}=e^{-i\omega_R t}e^{\omega_I t} $.
The real part $\omega_R$ clearly represents the oscillation frequency of the perturbation mode, whereas the imaginary part $\omega_I$ indicates the rate of exponential growth or decay of the perturbation. 
If $\omega_I$ is negative, it indicates that the perturbation decays over time, suggesting that the mode is stable; conversely, a positive $\omega_I$ signifies an unstable mode, under which the black hole spacetime might undergo substantial changes.

\section{Numerical Methods and Quasinormal Modes}\label{sec3}
To determine the frequencies of quasinormal modes, one must impose physically meaningful boundary conditions on the solutions of the wave equation in black hole spacetimes. 
These conditions dictate the asymptotic behavior of the scalar field, which varies depending on the spacetime geometry. 
In general, the field's asymptotic form is expressed as:
\begin{equation} \label{eq:boundary_conditions}
\psi(r) \sim
\begin{cases}
e^{G(\omega)r_*}, & r_* \to -\infty, \\
r^\alpha e^{K(\omega) r_*}, & r_* \to \infty,
\end{cases}
\end{equation}
where $ G(\omega) $ and $ K(\omega) $ are functions governed by the spacetime's asymptotic properties, specifically the effective potential near the boundaries, and $ \alpha $ reflects the physical system's characteristics \cite{Dolan:2007mj,Correa:2024xki}. 
In Einstein—Bumblebee black holes with massive fields, for example, we set $ G(\omega) = -i\sqrt{1+\ell} \omega $ and $ K(\omega) = \pm \sqrt{(1+\ell)(\omega^2 - \mu^2)} $. 
Here, solutions with $ \text{Re}[K(\omega)] > 0 $ represent quasinormal modes with outgoing behavior at infinity, while those with $ \text{Re}[K(\omega)] < 0 $ indicate exponentially decaying quasibound states, the latter being outside this study's scope \cite{Dolan:2007mj}. 
These distinct spectra coexist, each influencing the field's evolution differently, with quasibound states notably linked to superradiant instabilities. 
By contrast, in Einstein-Bumblebee de Sitter spacetime, where the effective potential diminishes at both the event and cosmological horizons, we adopt $ G(\omega) = -i\sqrt{1+\ell} \omega $ and $ K(\omega) = i\sqrt{1+\ell} \omega $, yielding oscillatory solutions at both limits:
\begin{equation}
\psi(r) \sim
\begin{cases}
e^{-i\sqrt{1+\ell}\omega r_*}, & r_* \to -\infty, \\
 e^{ i\sqrt{1+\ell}\omega r_*}, & r_* \to \infty,
\end{cases}
\end{equation}
irrespective of the field’s mass, thus eliminating the quasibound spectrum. 
In what follows, we will explore three distinct approaches to calculate the quasinormal frequencies in Einstein—Bumblebee de Sitter black holes.

\subsection{Matrix Method}
The Matrix Method (MM) offers a versatile approach to solving eigenvalue problems by employing an interpolation strategy that avoids reliance on a structured grid. This technique transforms the perturbation equations, together with their boundary constraints, into a homogeneous matrix equation through the discretization of the underlying linear partial differential equation \cite{Lin2017}. 
At the core of this method, the unknown eigenfunction is approximated using Taylor series expansions, constructed up to the $N$th order around $N$ selected points within the domain. These points, while capable of being irregularly spaced to optimize precision versus computational cost, are in this instance chosen to form a uniform grid over $[0,1]$. Solving the resulting system of linear algebraic equations yields the desired eigenvalue.

To simplify the radial domain, a change of variable is introduced:
\begin{equation}
x=\frac{r-r_h}{r_c-r_h},
\end{equation}
mapping the interval $r \in [r_h, r_c]$, which represents the outer communication region of a black hole, onto $x \in [0, 1]$. 
The function $\chi(x)$ is defined to vanish at the boundaries, i.e., $\chi(0) = 0$ and $\chi(1) = 0$, and is linked to the wave function $\psi_l^{\pm}(x)$ via:
\begin{align}\label{psi}
\psi_l^{\pm}(x)=x(1-x)(1-x)^{i\sqrt{(1+\ell)}\omega\eta_c}x^{- i\sqrt{1+\ell}\omega \eta_h} \chi(x).
\end{align}
The perturbation equations are then reexpressed in a compact differential form:
\begin{equation}\label{qicieq}
\tau_0^{\pm}(x)\chi''(x)+\lambda_0^{\pm}(x)\chi'(x)+s_0^{\pm}(x)\chi(x)=0,
\end{equation}
where the coefficients $\tau_0^{\pm}(x)$, $\lambda_0^{\pm}(x)$, and $s_0^{\pm}(x)$ are numerically evaluated based on the effective potentials, the coupling parameter $\xi$, and the black hole horizons $r_h$ and $r_c$. These horizons depend on the mass $M$, the effective cosmological constant $\Lambda_e$, and the multipole number $\ell$.

For numerical solution, the differential equation is discretized across a uniform set of grid points in $[0,1]$. Around each point, $\chi(x)$ is expanded via Taylor series, enabling the construction of differential matrices. This process converts the differential equation into a matrix system:
\begin{align}\label{algeq}
\left(M^{\pm}_0+\omega M_1^{\pm}\right)\chi(x)=0,	
\end{align}
where $\mathbf{\chi}$ is a vector containing the values of $\chi(x)$ at the grid points, and $M_0^{\pm}$ and $M_1^{\pm}$ are matrices built from the coefficients and differential operators. 
The QNM frequencies $\omega$ are determined by finding the values that satisfy:
\begin{equation}
|M^{\pm}_0+\omega M_1^{\pm} |=0.
\end{equation}
A distinguishing feature of the MM is its adaptability: the placement of interpolation points can be tailored to enhance accuracy in critical regions, though here a uniform distribution is adopted for simplicity. This flexibility makes the method particularly effective for balancing computational efficiency with the precision required in complex physical systems like black hole perturbation analyses.

\subsection{Wenzel—Kramers—Brillouin (WKB) Approximation Method}
Alternatively, to verify the results of the matrix method, we can employ the WKB method to calculate the quasi-normal modes when $\mu = 0$.
In 1926, Wenzel, Kramers, and Brillouin introduced a semiclassical approximation method for solving the Schrödinger equation.  
In the field of quantum mechanics, this method is extensively used for semiclassical calculations, such as calculating the penetration rates of bound state potential barriers.  
The WKB approximation integrates elements of both classical and quantum theories, offering an effective tool for addressing challenges that are difficult to resolve through purely classical or quantum approaches.  
This is particularly advantageous in dealing with potential barrier penetration and quantum tunneling effects.  

For a general effective potential $V_\text{eff}(r)$, the formula for the sixth-order WKB approximation is as follows~\cite{Iyer1987,Konoplya2003}:
\begin{equation}
i\frac{\omega^2-V_0}{\sqrt{-2V''_0}}-\Lambda_2-\Lambda_3-\Lambda_4-\Lambda_5-\Lambda_6=n+\frac{1}{2},
\end{equation}
where $V_0=V_{r_*=r_{max}}$ is the peak position of the effective potential, 
and $V''_0=\frac{d^2}{dr^2_*}V_{r_*=r_{max}}$ denotes the second derivative of the potential with respect to the tortoise coordinate at $r_{max}$, 
which corresponds to the maximum value of the potential $V$.  
$n$ is the overtone number, and $\Lambda_i$ represent the correction coefficients in the WKB method.  
The order of these corrections determines the precision of our numerical results.  
Owing to their complexity, the methodology and the specific forms of the coefficients are described in reference~\cite{Konoplya2003}.

\subsection{Quasinormal Mode Numerical Results}

Scalar modes exist for any angular quantum number $L$, with particular attention given to three modes: $L=0$ for the monopole mode, $L=1$ for the dipole mode, and $L=2$ for the quadrupole mode. 
The impact of the Lorentz violation parameter, $\ell$, on the frequencies of these quasinormal modes is detailed for each mode.
\begin{figure}[t!]
\centering
\includegraphics[width=0.3\textwidth]{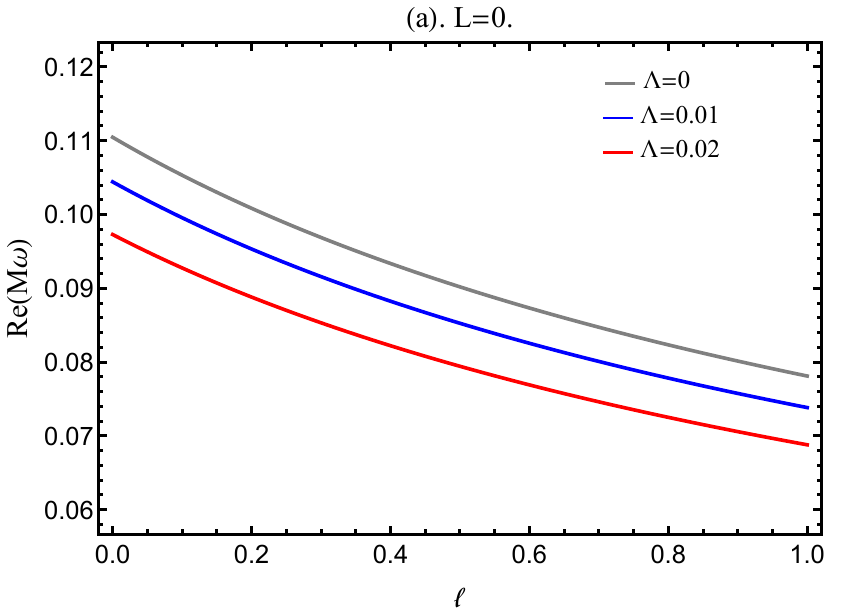} 
\includegraphics[width=0.3\textwidth]{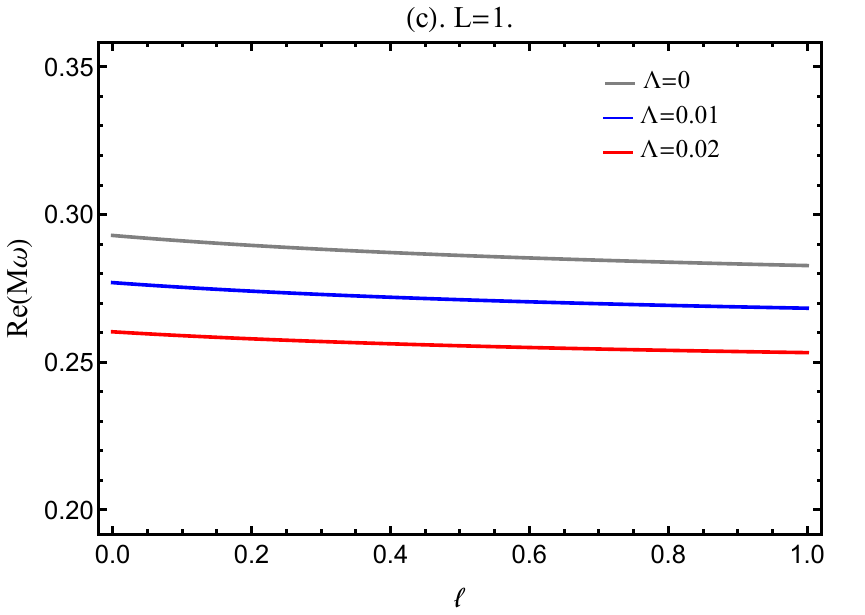} 
\includegraphics[width=0.3\textwidth]{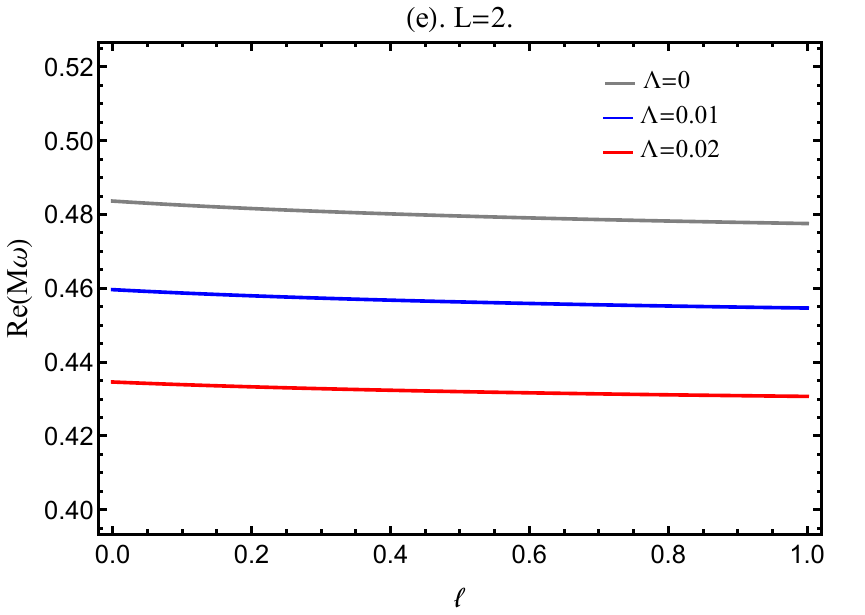} 
\includegraphics[width=0.3\textwidth]{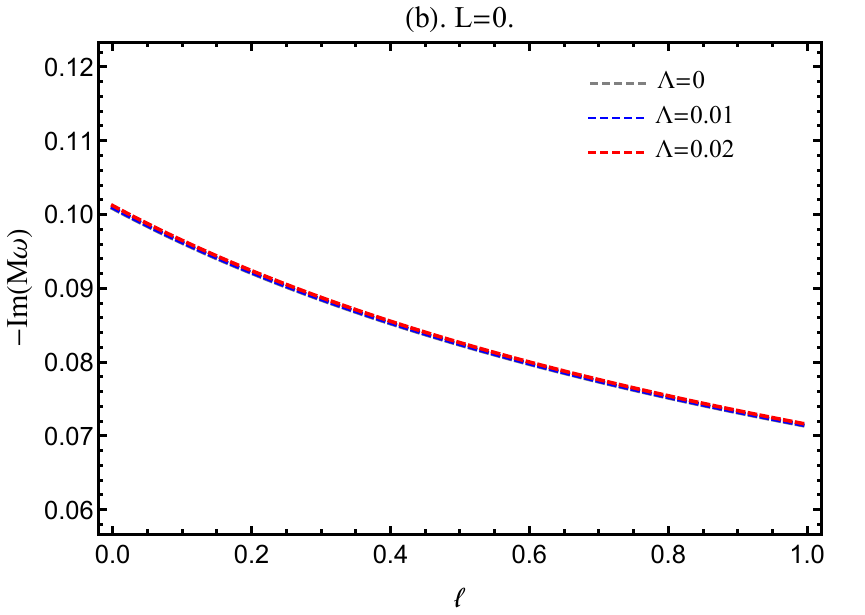} 
\includegraphics[width=0.3\textwidth]{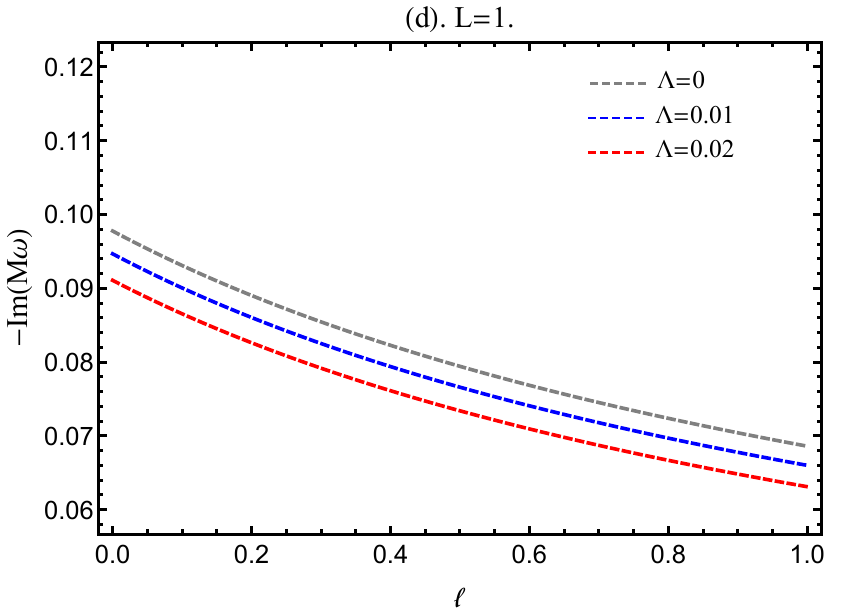} 
\includegraphics[width=0.3\textwidth]{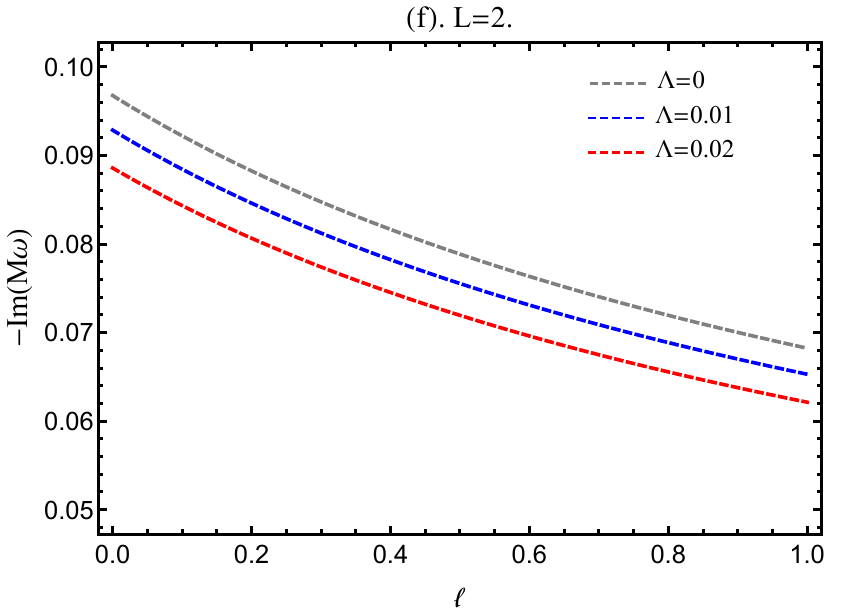} 
\caption{Complex massless scalar frequencies for the $n=0$ mode for varying values of $\ell$.}
\label{fig1}
\end{figure}

Figs. \ref{fig1}(a)—\ref{fig1}(f) show the QNMs of a massless scalar field using the WKB method.
As the Lorentz violation parameter $\ell$ increases, both the real and imaginary parts of the quasinormal modes in the monopole modes tend to decrease. 
In dipole and quadrupole modes, changes in the real part of the quasinormal modes are not significant, but the imaginary parts decrease rapidly. 
Additionally, when the cosmological constant $\Lambda$ increases, the overall behavior of the quasinormal modes tends to decrease. 
The impacts on the dipole and quadrupole modes are similar in trend but different in magnitude. 
Anomalous behavior is observed in the monopole mode, where the cosmological constant $\Lambda$ seems to have a negligible effect on the imaginary part. 
The validity of these results will be verified in the next section through fitting dynamical waveforms.

\begin{figure}[t!]
\centering
\includegraphics[width=0.3\textwidth]{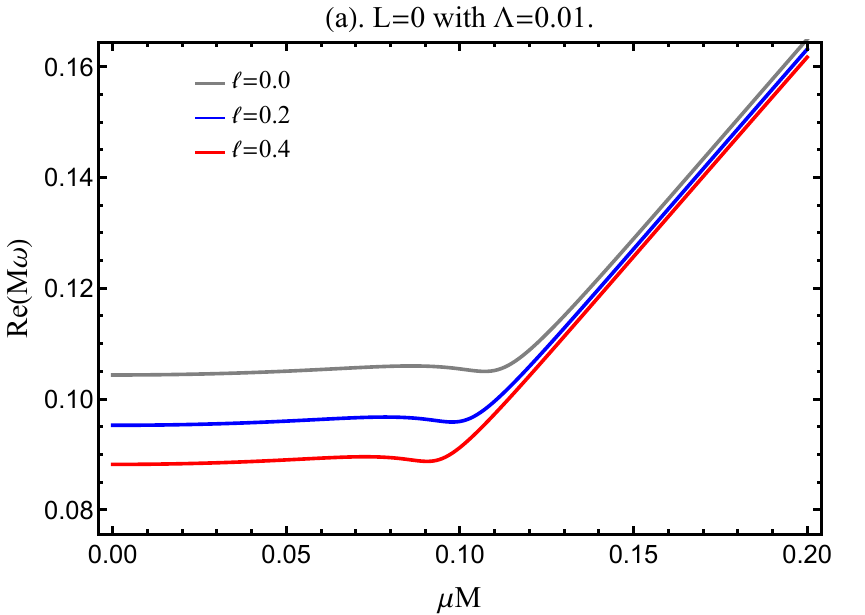} 
\includegraphics[width=0.3\textwidth]{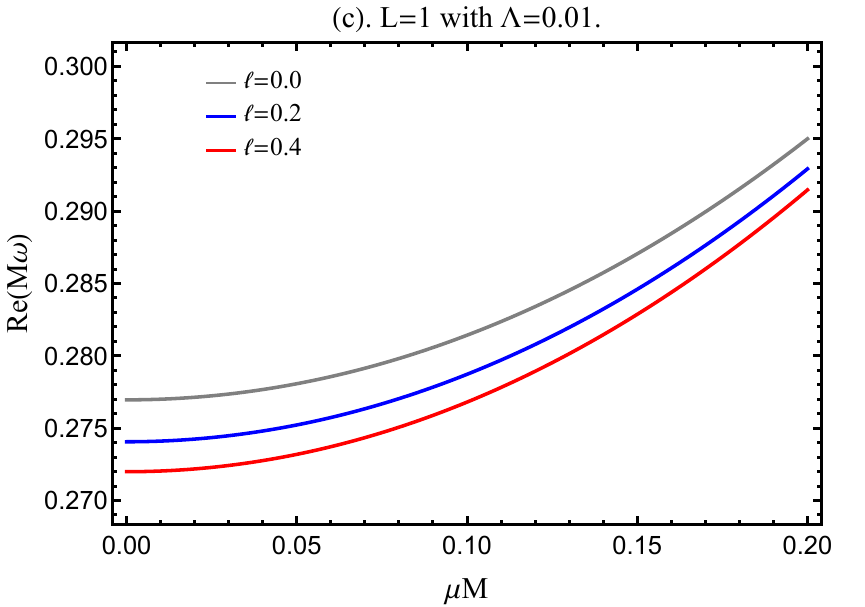} 
\includegraphics[width=0.3\textwidth]{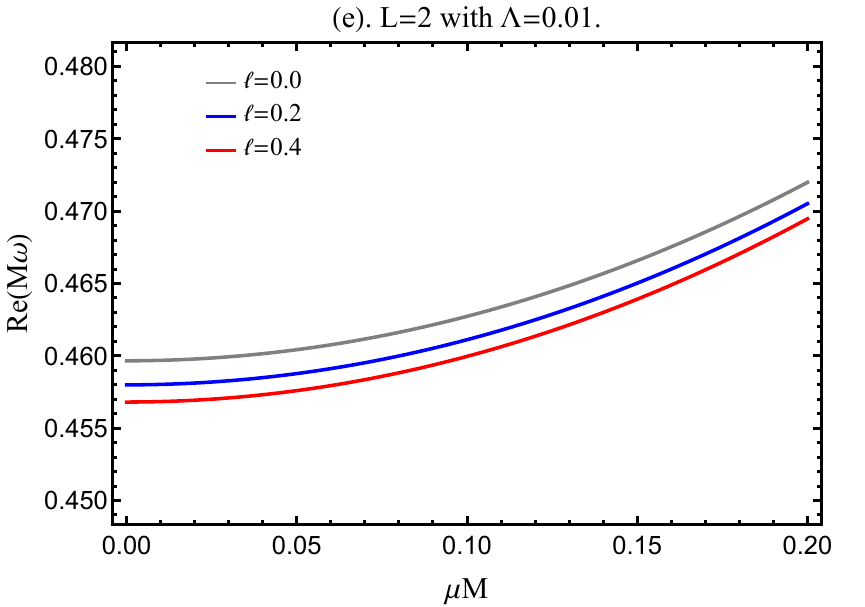} 
\includegraphics[width=0.3\textwidth]{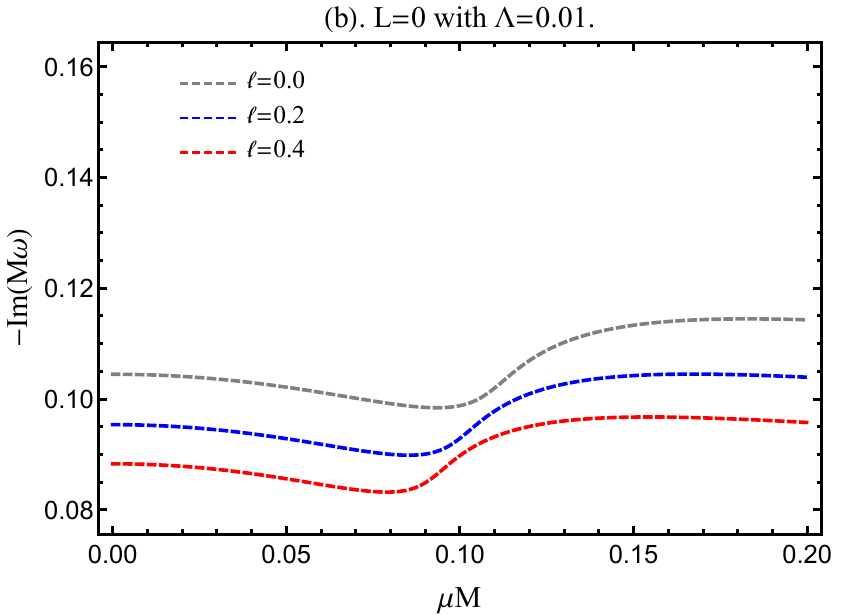} 
\includegraphics[width=0.3\textwidth]{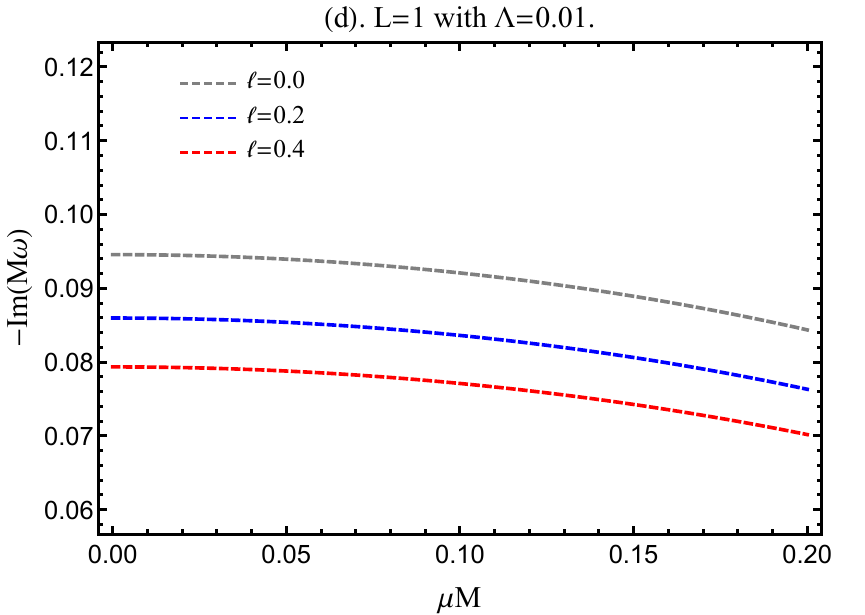} 
\includegraphics[width=0.3\textwidth]{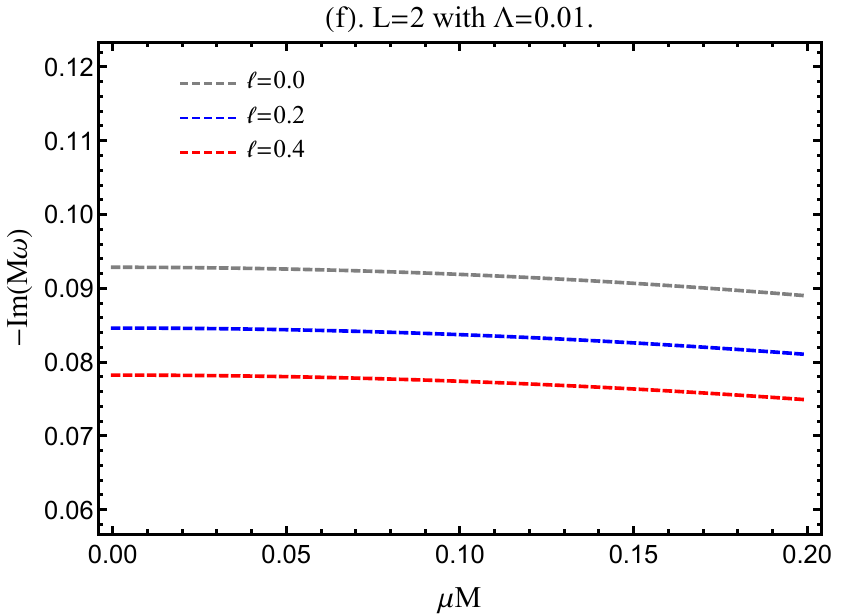} 
\caption{Complex massive scalar frequencies for the $n=0$ mode for varying values of $\mu M$.}
\label{fig1n}
\end{figure}
Figs. \ref{fig1n}(a)—\ref{fig1n}(f) show the variation of the real and imaginary parts of QNMs with increasing field mass under different Lorentz-violating parameters in the presence of a cosmological constant using the MM method with $ N=30 $.
For the dipole and quadrupole modes, the effect of the field mass on the frequency is monotonic, with the real part increasing and the imaginary part decreasing as the field mass increases. It is noteworthy that all parameters discussed in this paper reduce the imaginary part of QNMs; however, the Lorentz-violating parameters result in a slower decrease, as observed in the figures, while the field mass leads to a more rapid decrease.
For the monopole mode, the introduction of the field mass leads to significantly different results. Within the range where the field mass satisfies $ \mu M < 0.1 $, it appears to have no effect on QNMs, whereas in the large mass regime, its influence resembles that of other modes. 
This is because the field mass alters the boundary behavior of the scalar field’s effective potential at infinity, which is consistent with the effect of the cosmological constant. 
The monopole mode, lacking angular quantum numbers, is governed by the cosmological constant in terms of the effective potential's behavior, which accounts for the aforementioned phenomenon.

\section{Dynamical Evolution}\label{sec4}

In a finite time domain, the numerical evolution of an initial wave packet governed by the time-dependent Schrödinger equation~\eqref{eq:schrodinger} can be considered~\cite{Konoplya2003}. 
Using the Eddington-Finkelstein coordinates, the following coordinate transformation is applied to the equation \cite{Santos:2015gja}:
\begin{equation}
u = t_* - r_*, \quad v = t_* + r_*,
\end{equation}
where the time-dependent Schrödinger equation can be rewritten as:
\begin{equation}\label{eq:wave_uv}
4\frac{\pp^2\psi(u,v)}{\pp_u\pp_v}-V(u,v)\psi(u,v)=0.
\end{equation}

Using light-cone coordinates simplifies the analysis of the wave function~\cite{Das2021}, as the equation adopts a simpler form in these coordinates. 
In particular, the equation transforms into a wave equation in the variables $u$ and $v$, which can be addressed via standard methods such as the finite difference method (FDM)~\cite{Abdalla:2010nq,Zhu:2014sya,Lin:2022rtx}. 
The waveform of $\psi(u,v)$ is determined by applying the following initial conditions to equation~\eqref{eq:schrodinger}:
\begin{equation}\label{eq:initial_conditions}
\psi(u,0)=0,\quad \psi(0,v)=Exp\left[ -\frac{(v-v_c)^2}{2\gamma^2} \right],
\end{equation}
Here, $ \psi(0,v) $ represents a Gaussian wave packet with width $ \gamma $ at $ v_c $.
The observer's position is set at ten times the radius of the black hole's event horizon, defined as $r_0 = 20 M$. 
The dynamical evolution waveform is subsequently obtained by numerically solving the partial differential equation.

\begin{figure}[h!]
    \centering
    \includegraphics[width=0.27\textwidth]{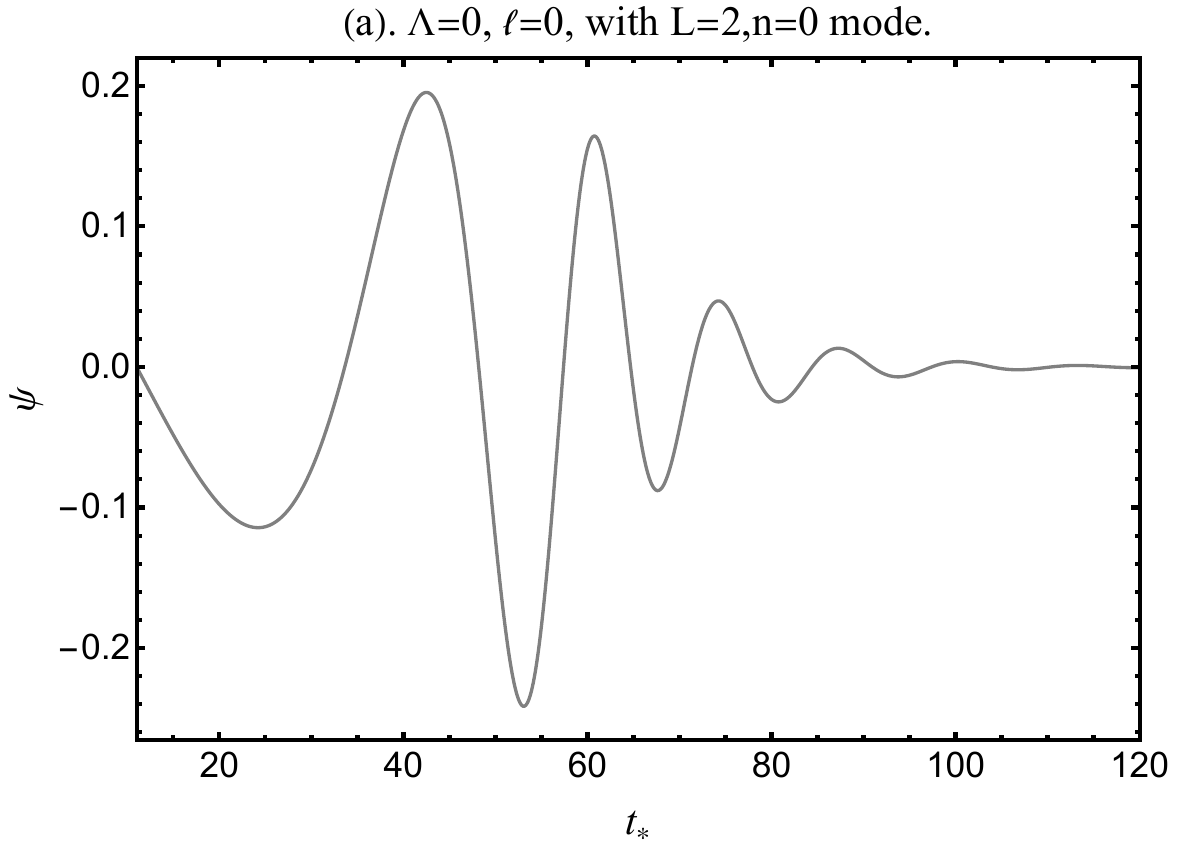} 
    \includegraphics[width=0.27\textwidth]{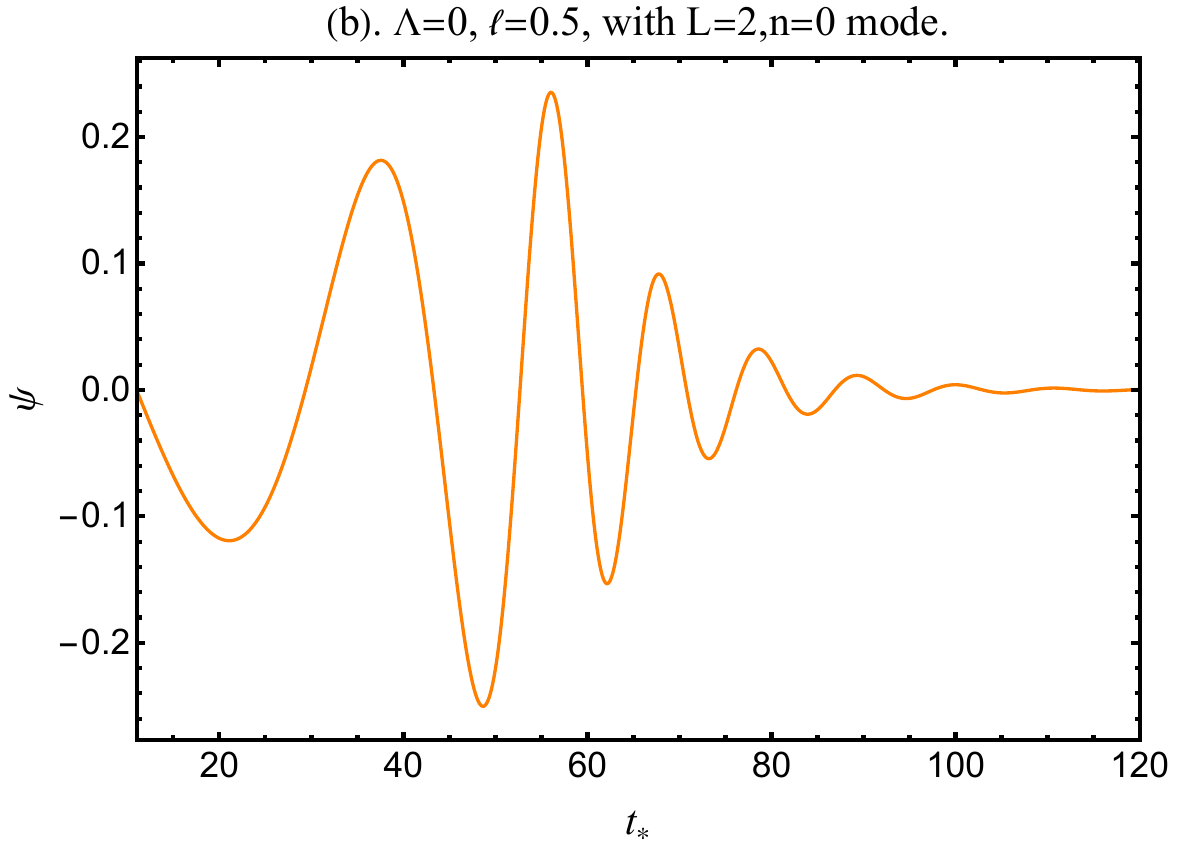} 
    \includegraphics[width=0.27\textwidth]{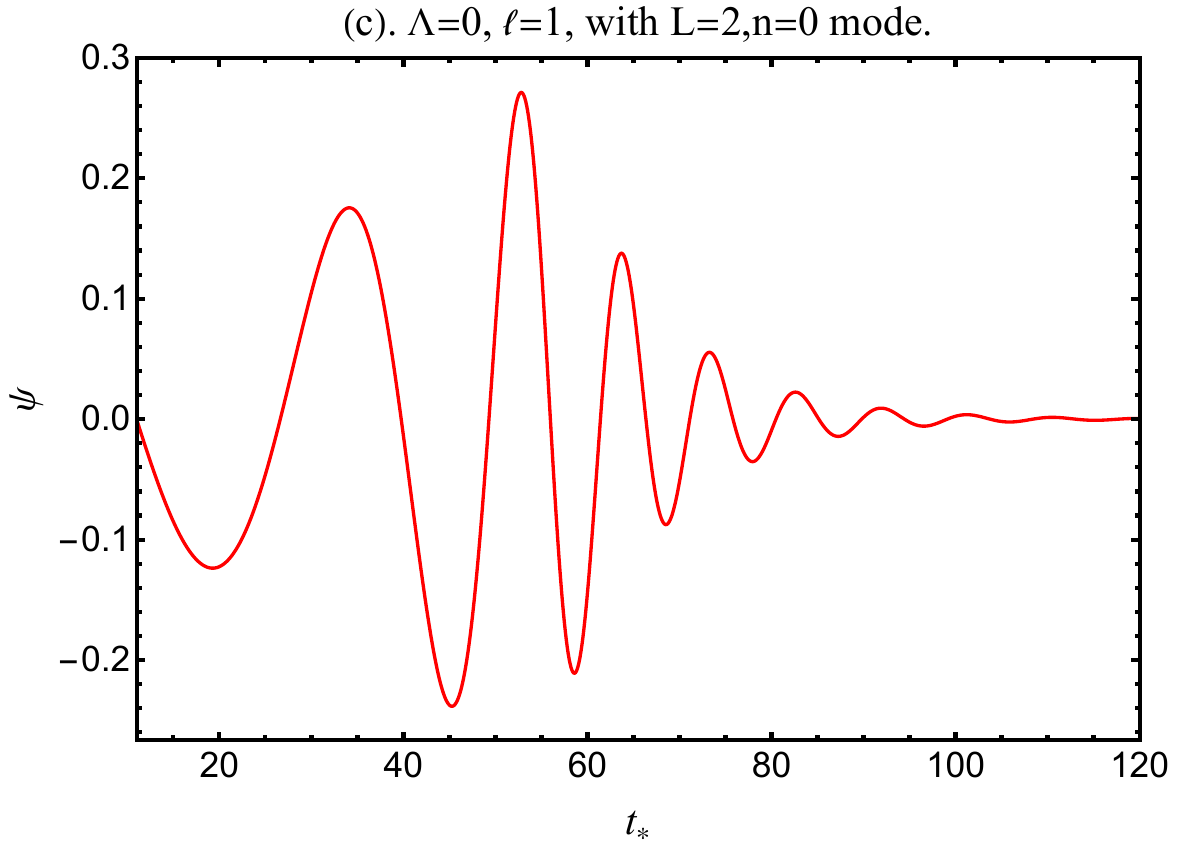} 
    \includegraphics[width=0.27\textwidth]{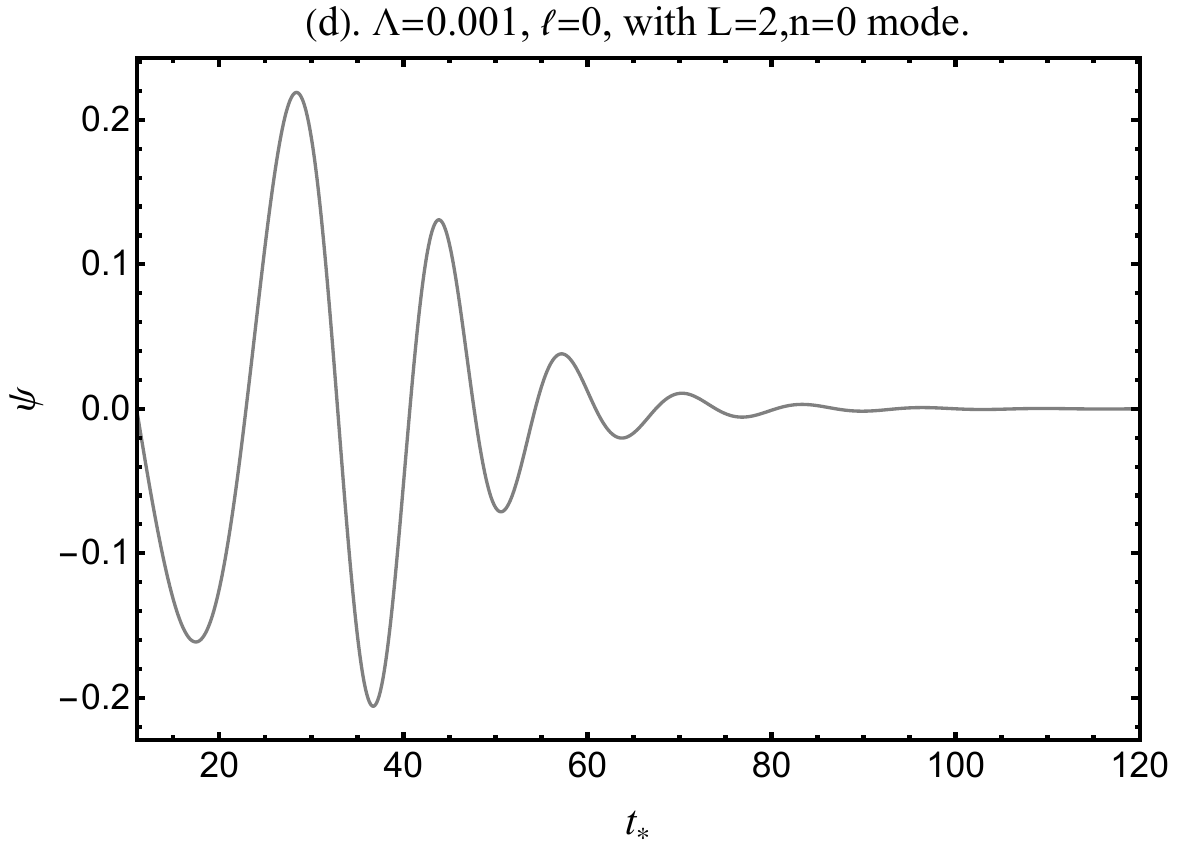} 
    \includegraphics[width=0.27\textwidth]{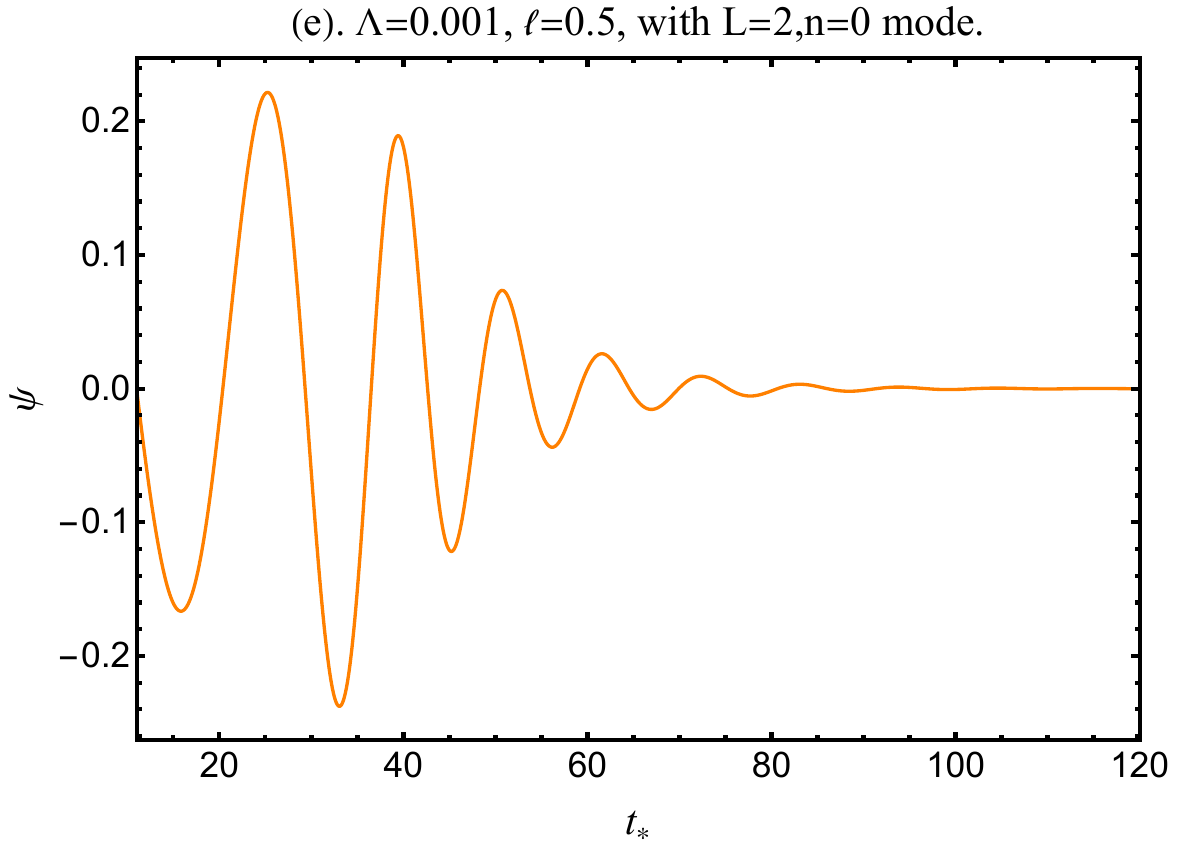} 
    \includegraphics[width=0.27\textwidth]{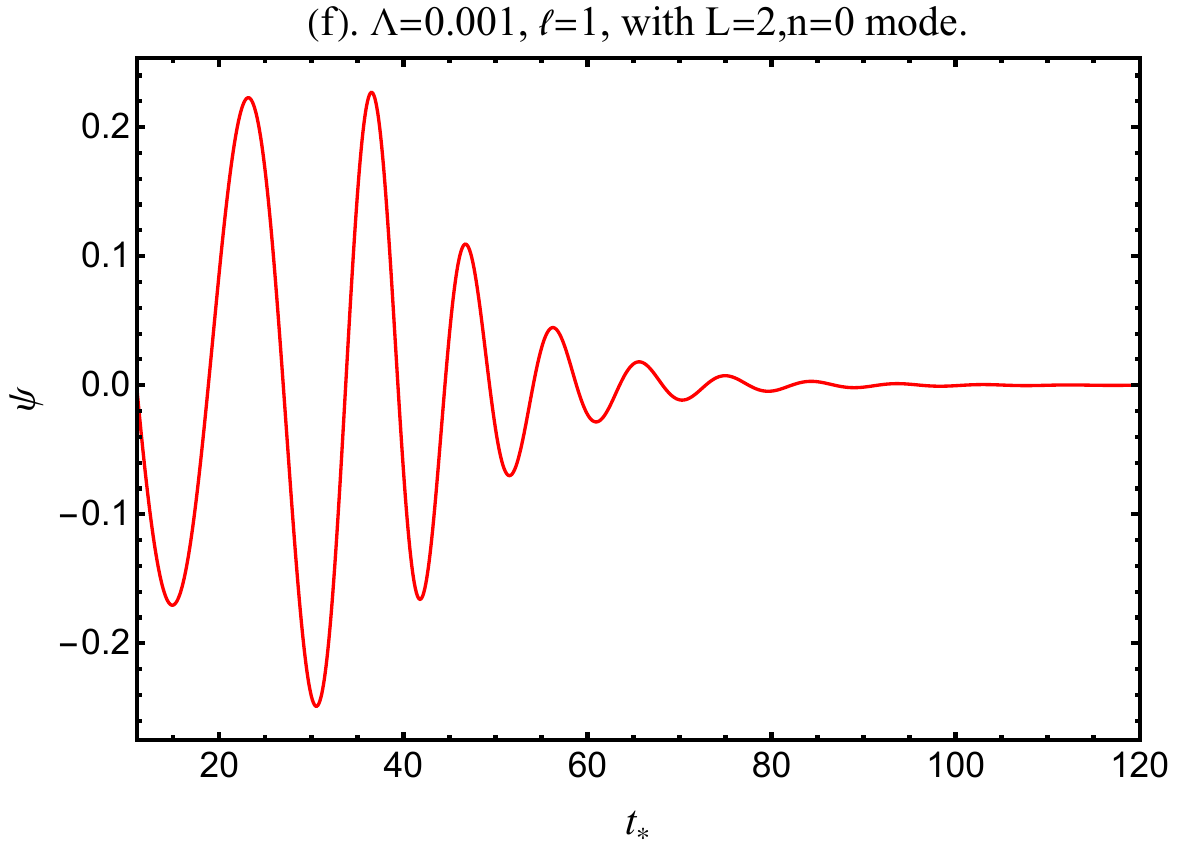} 
    \includegraphics[width=0.27\textwidth]{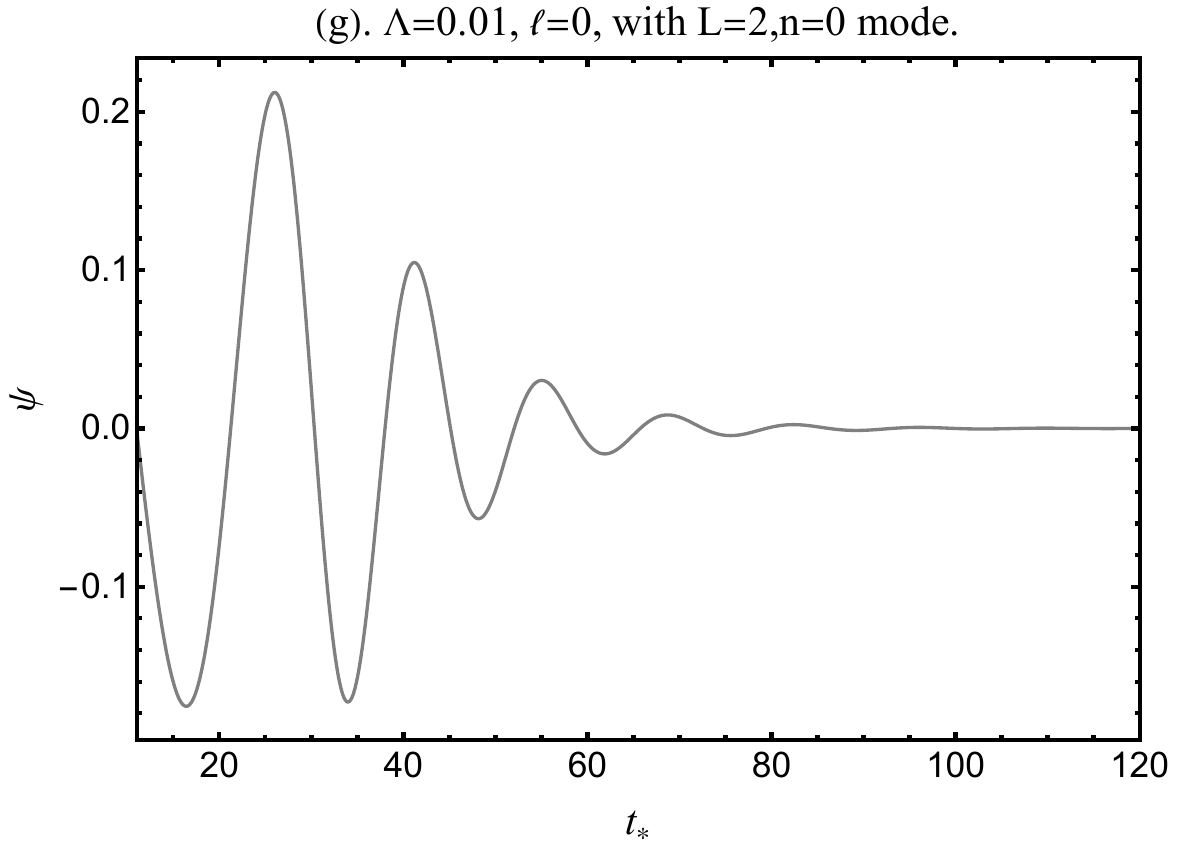} 
    \includegraphics[width=0.27\textwidth]{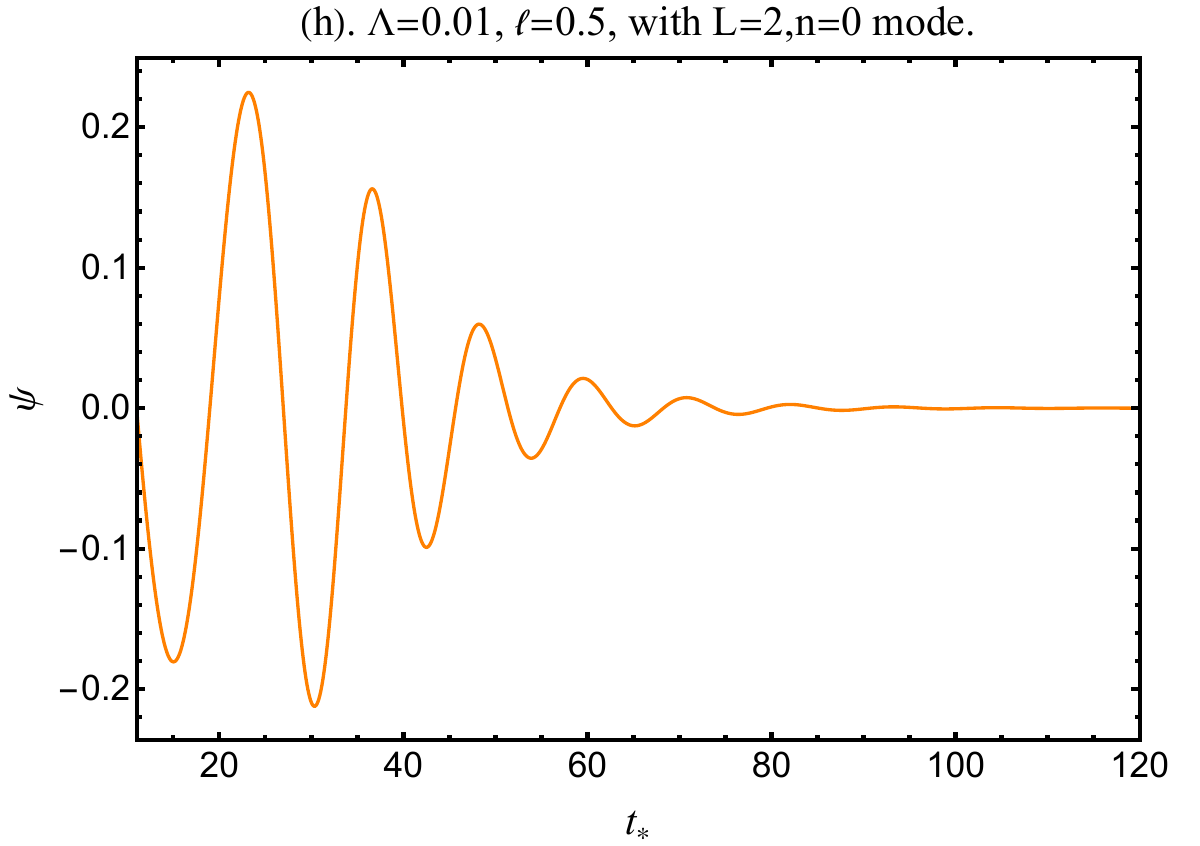} 
    \includegraphics[width=0.27\textwidth]{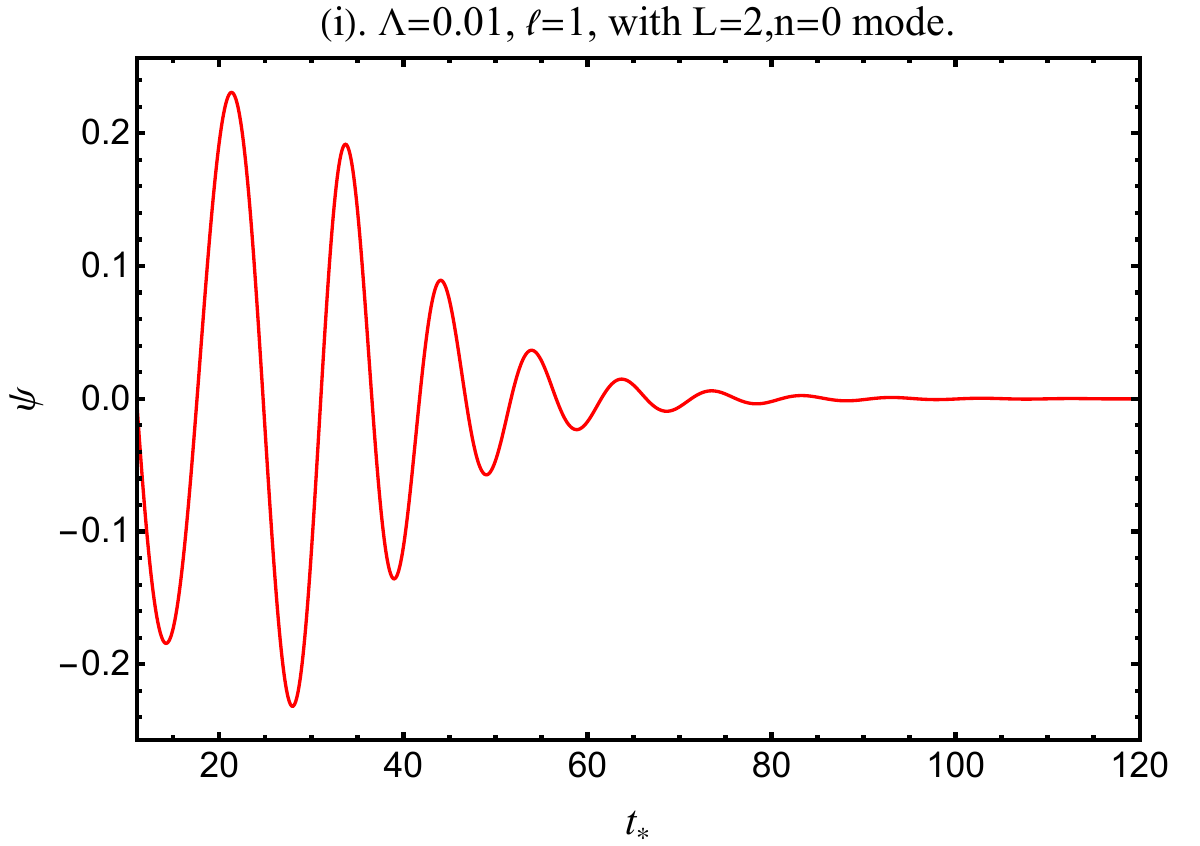} 
    \caption{The time evolution of the wave function $\psi$ corresponds to the scalar perturbation in the $L=2$ mode.}
    \label{fig3}
\end{figure}
\begin{figure}[h!]
    \centering
    \includegraphics[width=0.27\textwidth]{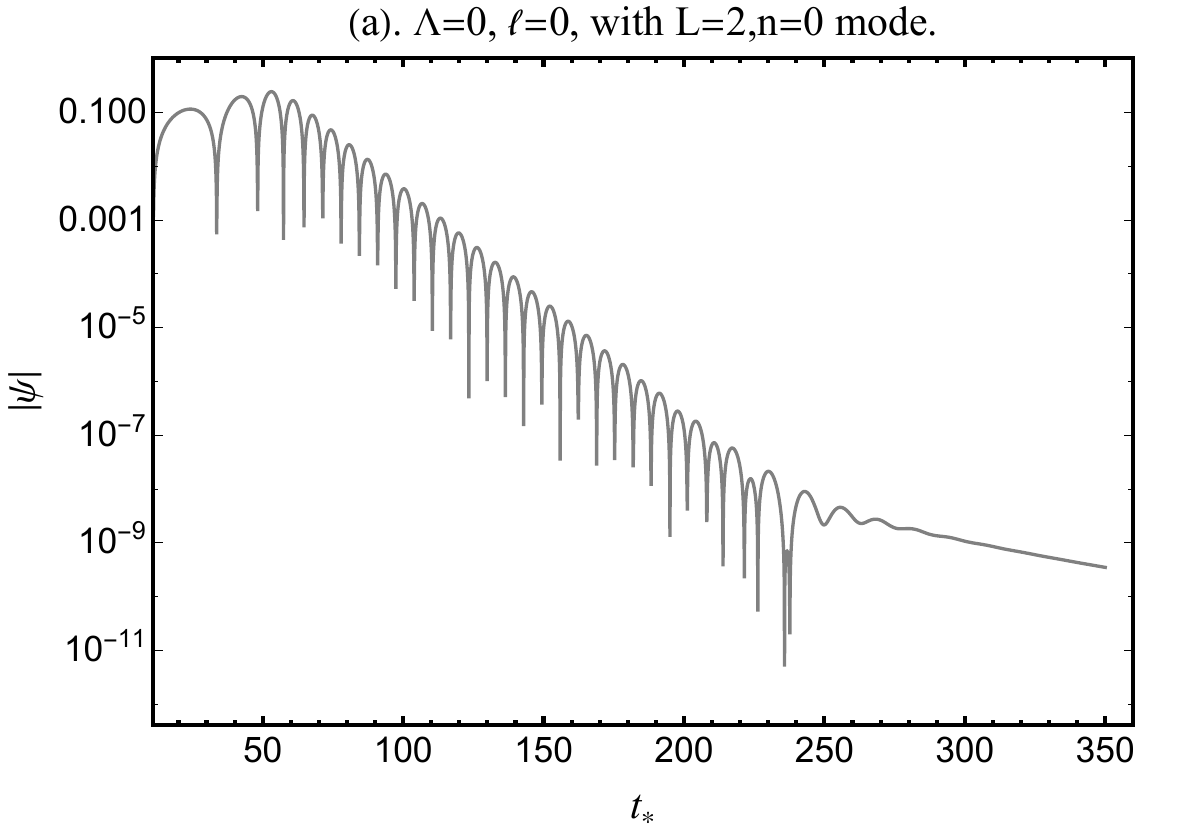} 
    \includegraphics[width=0.27\textwidth]{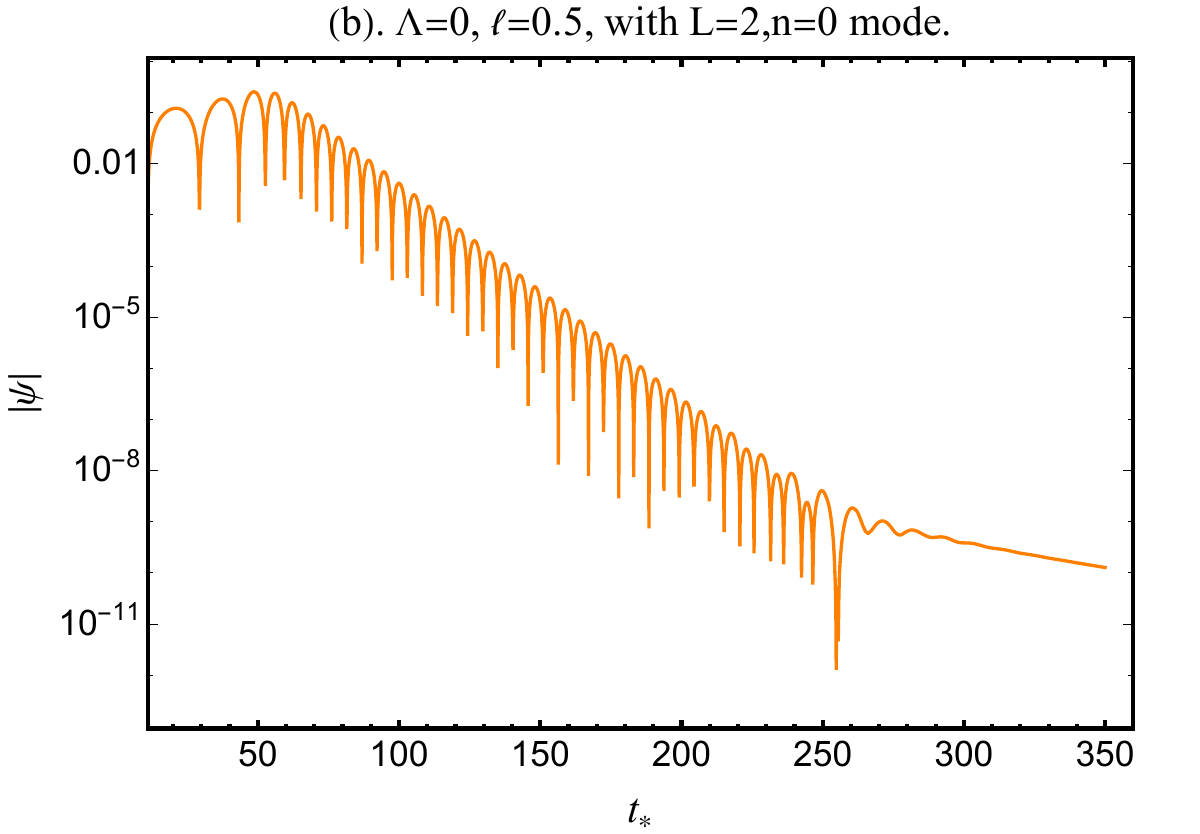} 
    \includegraphics[width=0.27\textwidth]{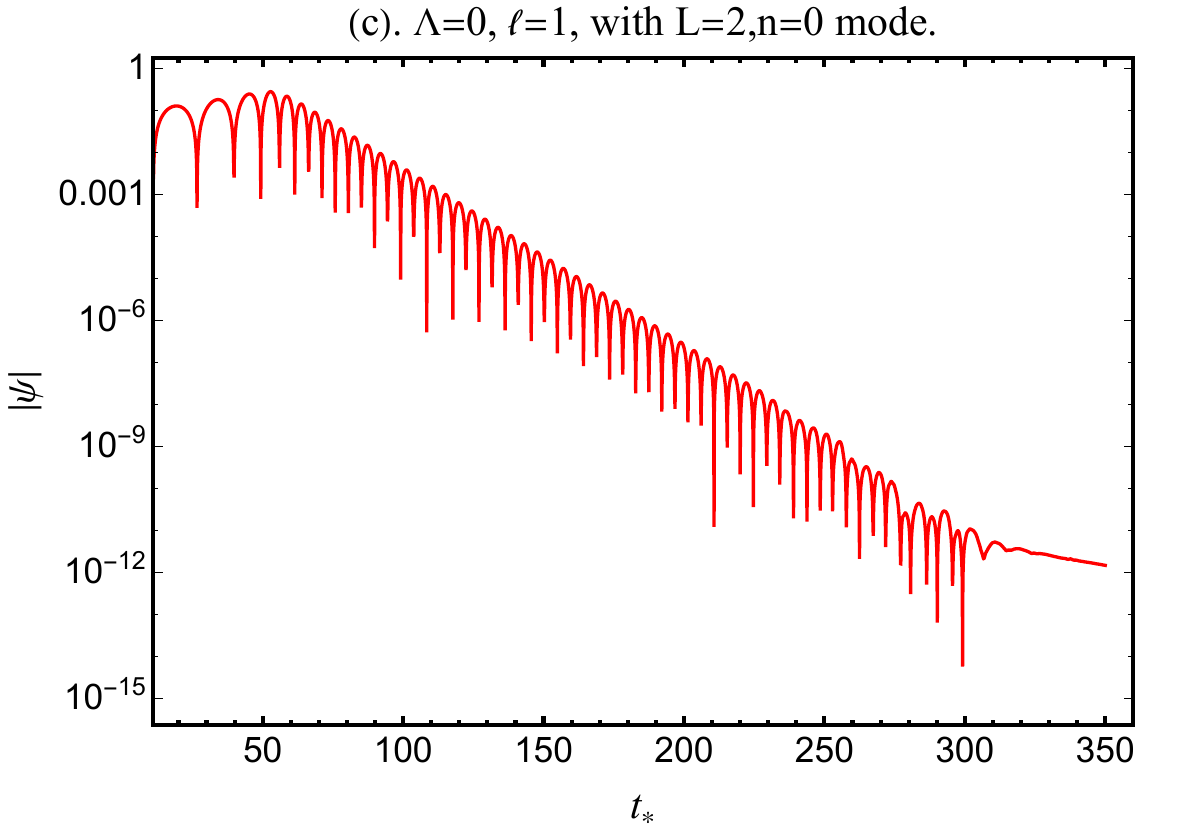} 
    \includegraphics[width=0.27\textwidth]{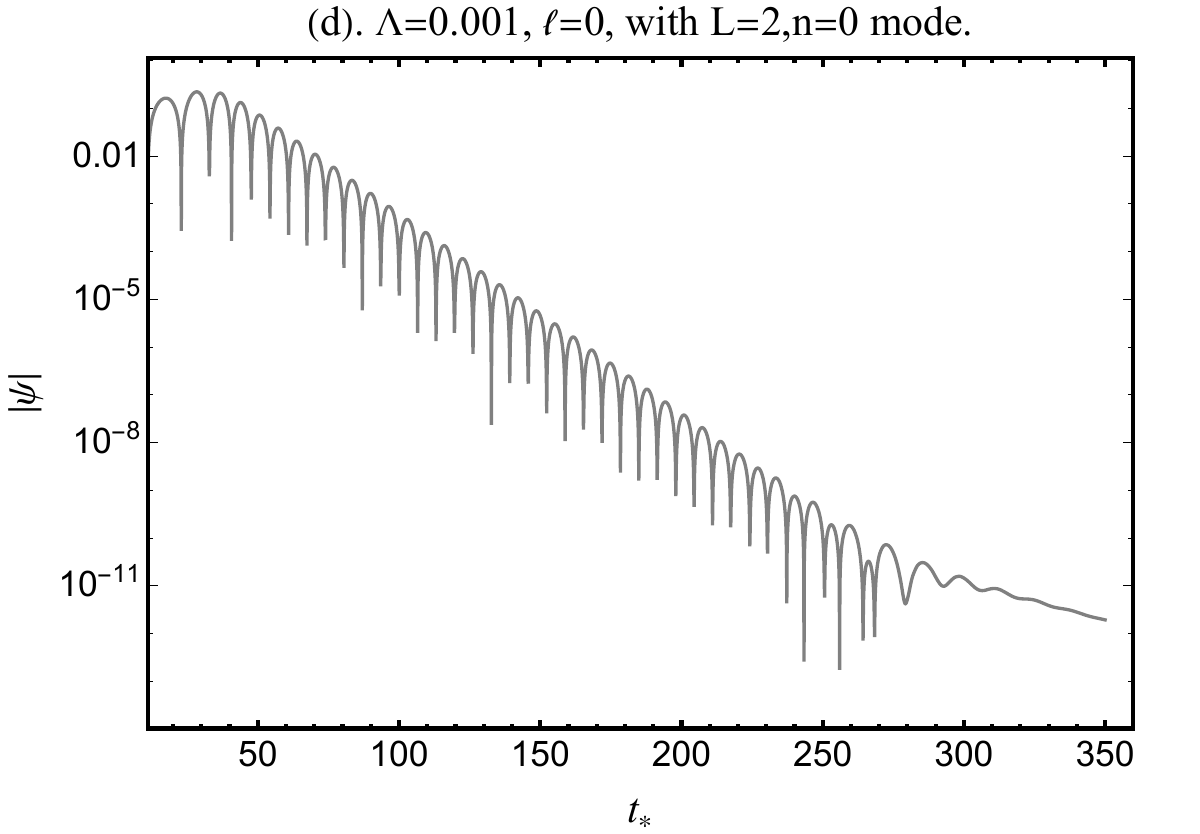} 
    \includegraphics[width=0.27\textwidth]{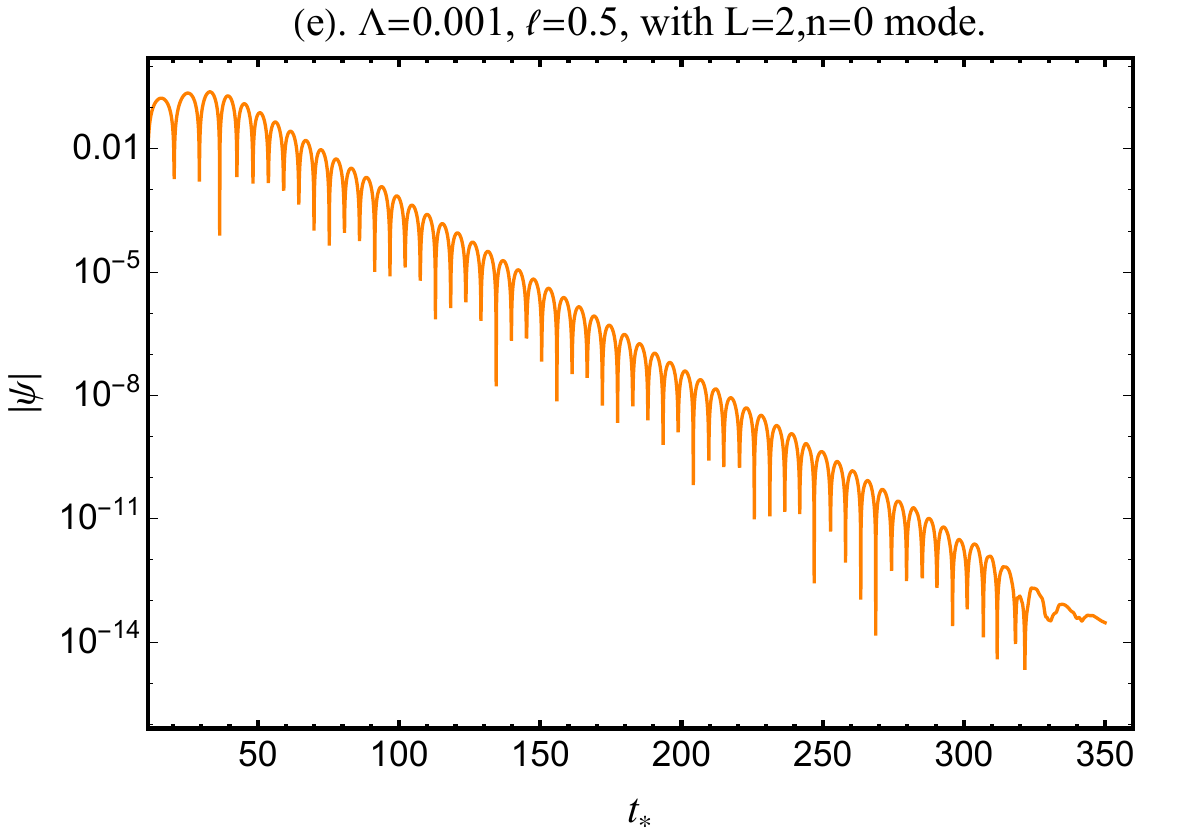} 
    \includegraphics[width=0.27\textwidth]{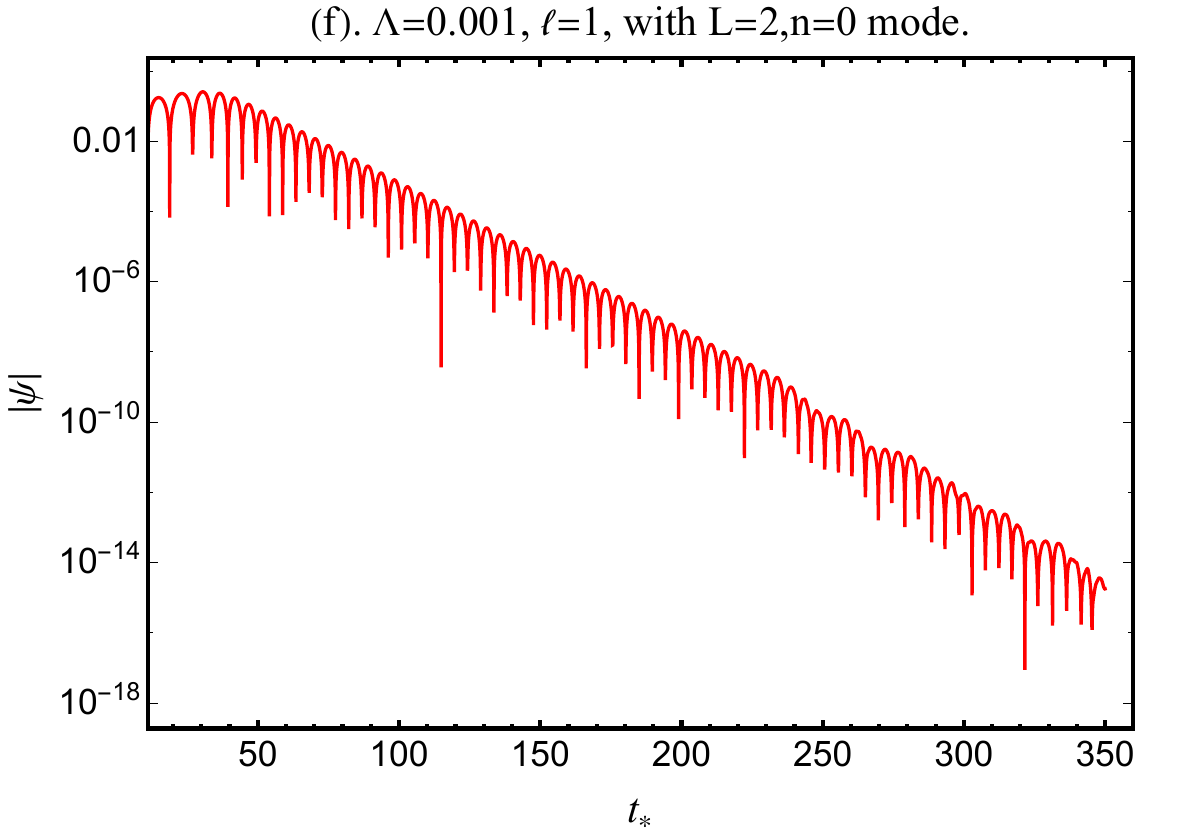} 
    \includegraphics[width=0.27\textwidth]{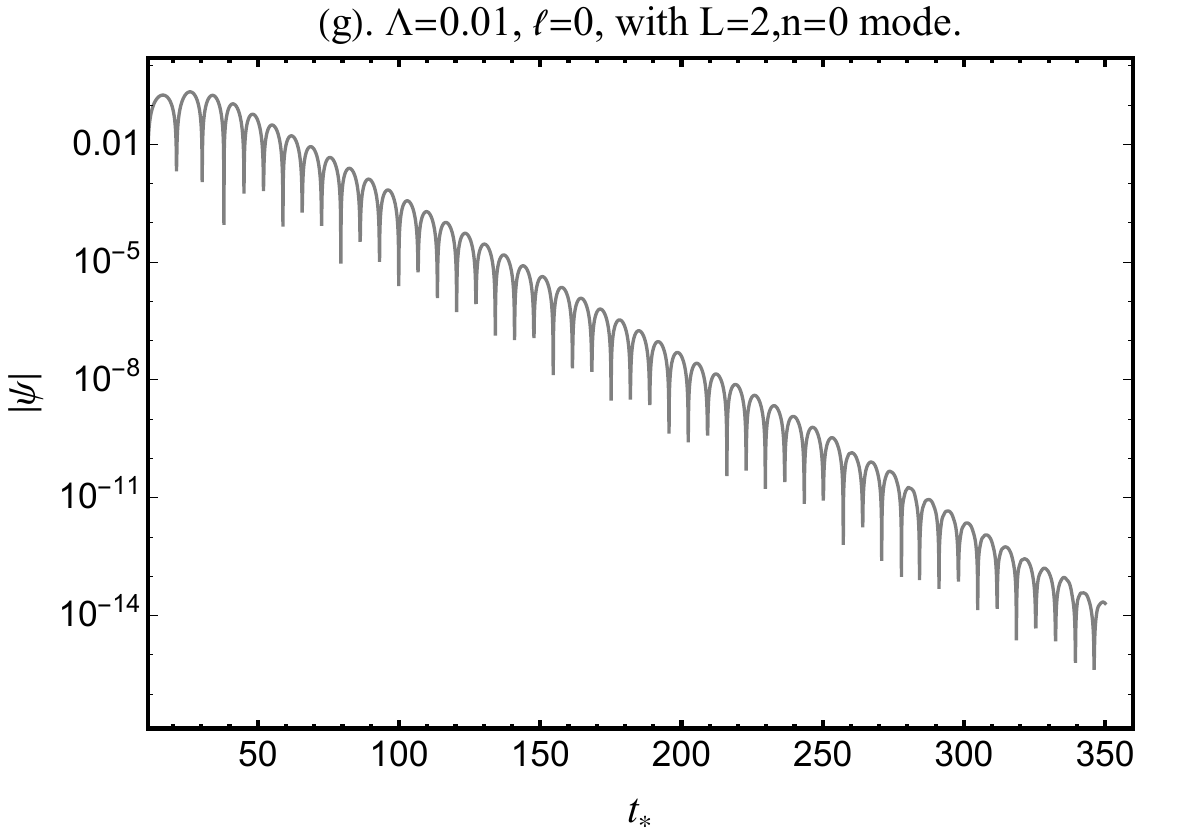} 
    \includegraphics[width=0.27\textwidth]{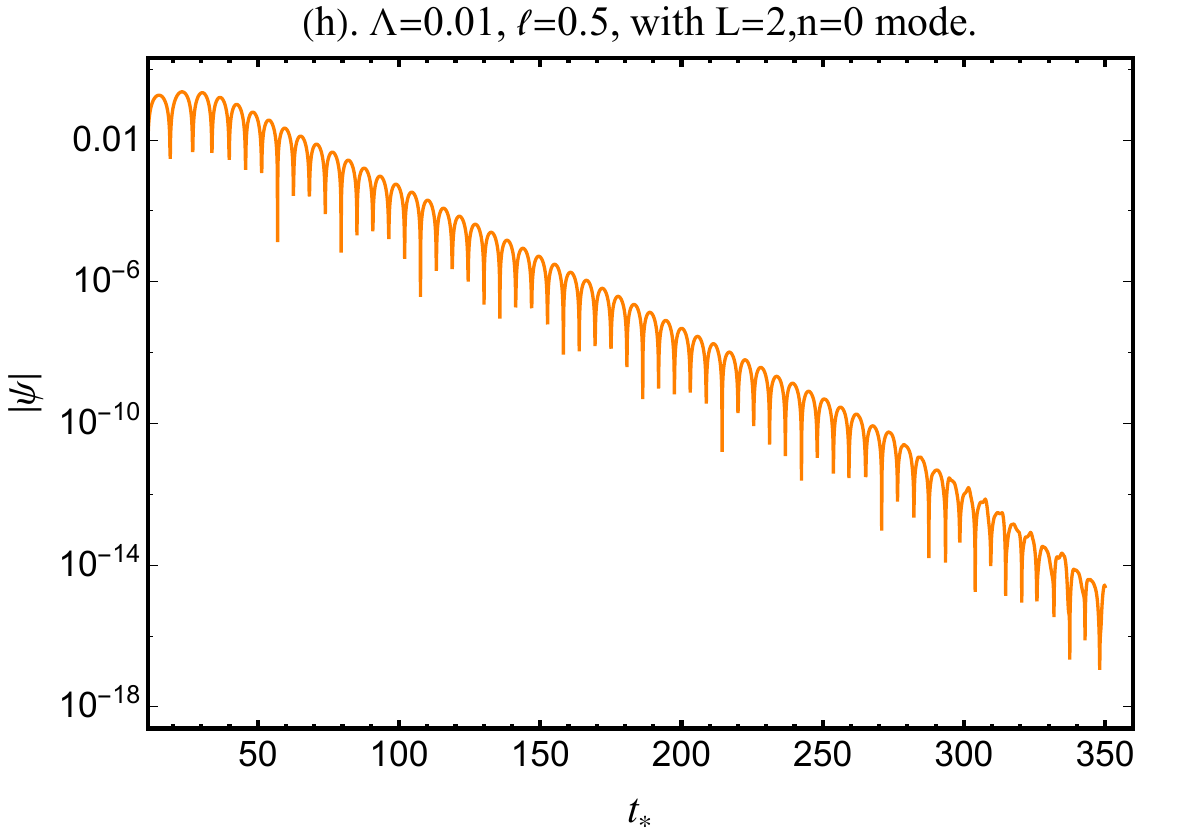} 
    \includegraphics[width=0.27\textwidth]{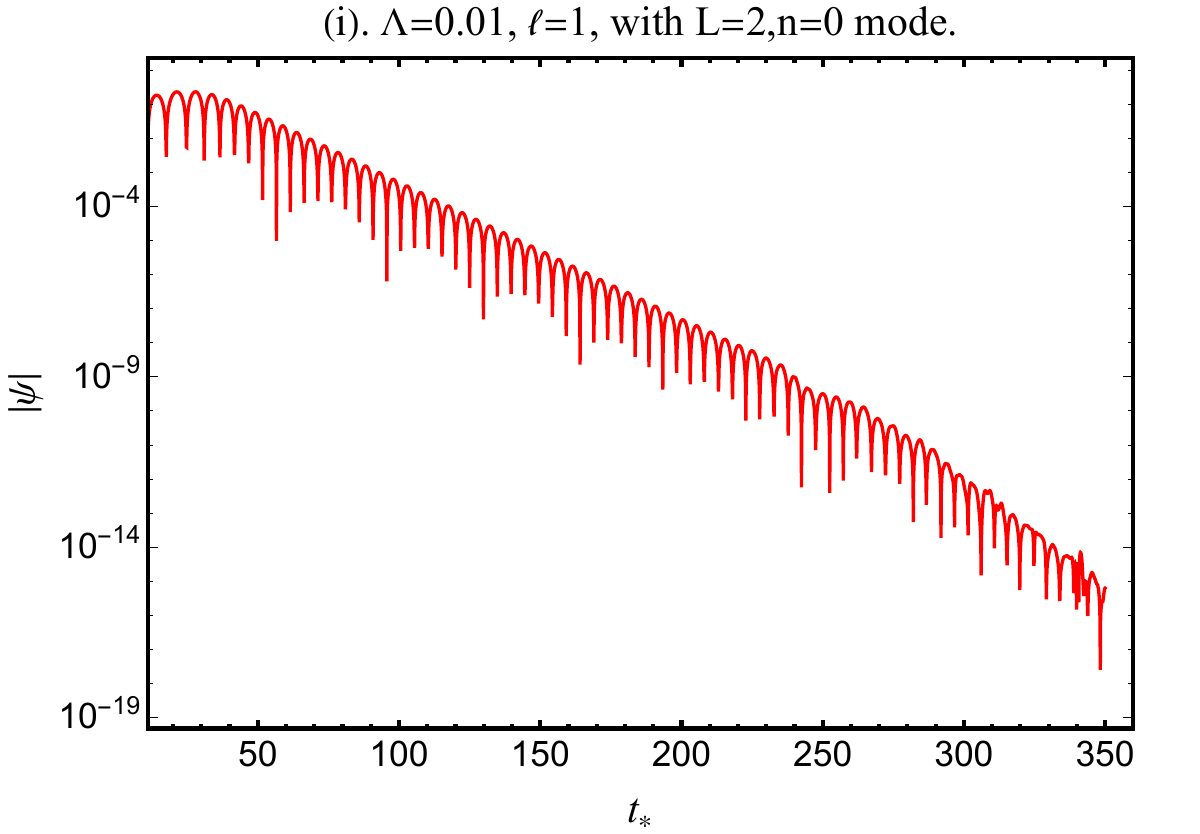} 
    \caption{The time evolution of the wave function $|\psi|$ corresponds to the scalar perturbation in the $L=2$ mode (Logarithmic version).}
    \label{fig2}
\end{figure}

Figs. \ref{fig3}(a)-\ref{fig3}(i) display the dynamical evolution waveforms of scalar field perturbations under various black hole parameter conditions. 
Here, we perform a coordinate transformation $(u,v) \rightarrow (t_*, r_*)$ on the result obtained for $ \psi(u,v) $ from equation (\ref{eq:wave_uv}) to restore it to the same spacetime coordinates used during the calculation of QNMs.
These waveforms correspond to cases where the cosmological constant $\Lambda$ takes values of $\Lambda=0$, $\Lambda=10^{-3}$, and $ \Lambda=10^{-2} $ and where the Lorentz violation parameter, $\ell$, is set to $\ell=0$, $\ell=0.5$, and $\ell=1$, under the quadrupole mode ($L=2$). 
We have added the logarithmic evolution plot of the scalar perturbation wave function for the $L=2$ mode in Figures~\ref{fig2}(a)-\ref{fig2}(i), where the linear decay of the amplitude $|\psi(t)|$ directly corresponds to the extraction of the imaginary part of the QNMs, $\omega_I$.
When the cosmological constant $\Lambda$ is zero, the images generally exhibit significant tailing phenomena. 
However, for moderate values of the cosmological constant, tailing only occurs when the Lorentz violation parameter $\ell$ is small. Under conditions of a relatively large cosmological constant, the tailing phenomenon completely disappears.
This finding indicates that in de Sitter spacetimes, the magnitude of the cosmological constant substantially dictates the occurrence of tailing, whereas Lorentz-symmetry breaking can make the tailing less pronounced. 
As the Lorentz violation parameter $\ell$ increases, the decay rate of the field decreases with a fixed cosmological constant, reflecting the physical manifestation of the decrease in the imaginary part of the quasinormal mode frequencies.

\begin{table*}[ht]
\centering
 \caption{A comparison is conducted between the matrix method, the WKB approximation, and the waveform fitting for the $L=2$ mode with different values of the cosmological constant and Lorentz-violating parameters.}
\label{tab:comparison}
\begin{tabular}{c c c c c}
\hline
\textbf{~~~~~~~~~~~~~}~~~& ~~~\textbf{~~~~~~~~~~~~~~~~~~~~~}~~~ &~~~~~~\textbf{Matrix Method}~~~~~~&~~~~~~\textbf{WKB}~~~~~~ & ~~~~~~\textbf{Fitting}~~~~~~\\
\hline
\multirow{3}{*}{$\Lambda = 0$}
  & $\ell = 0$  & $~~~~~~~~ - ~~~~~~~~~~~$~~~~  & $0.483642 - 0.096761\,i$~~~~   & $0.482891 - 0.096664\,i$   \\
  & $\ell = 0.5$& $~~~~~~~~ - ~~~~~~~~~~~$~~~~  & $0.479592 - 0.078854\,i$~~~~   & $0.479536 - 0.078892\,i$    \\
  & $\ell = 1$  & $~~~~~~~~ - ~~~~~~~~~~~$~~~~  & $0.477556 - 0.068236\,i$~~~~   & $0.475290 - 0.068328\,i$   \\
\hline
\multirow{3}{*}{$\Lambda = 10^{-3}$}
  & $\ell = 0$  & $0.482488 - 0.097261\,i$~~~~  & $0.481284 - 0.096391\,i$~~~~  & $0.481282 - 0.096459\,i$   \\
  & $\ell = 0.5$& $0.478058 - 0.080636\,i$~~~~  & $0.477306 - 0.078543\,i$~~~~  & $0.477337 - 0.078467\,i$   \\
  & $\ell = 1$  & $0.474743 - 0.071302\,i$~~~~  & $0.473508 - 0.067951\,i$~~~~  & $0.473536 - 0.068098\,i$   \\
\hline
\multirow{3}{*}{$\Lambda = 10^{-2}$}
  & $\ell = 0$  & $0.459638 - 0.092853\,i$~~~~  & $0.459636 - 0.092859\,i$~~~~  & $0.459627 - 0.092887\,i$   \\
  & $\ell = 0.5$& $0.456312 - 0.075539\,i$~~~~  & $0.456132 - 0.075540\,i$~~~~  & $0.456325 - 0.075619\,i$   \\
  & $\ell = 1$  & $0.454652 - 0.065296\,i$~~~~  & $0.454652 - 0.065296\,i$~~~~  & $0.454691 - 0.065451\,i$   \\
\hline
\end{tabular}
\end{table*}
Finally, we present the fitted results and compare them with those obtained using the Matrix Method and the WKB Approximation, as shown in Table \ref{tab:comparison}. 
Due to fundamental differences in the boundary conditions described earlier, when $\Lambda = 0$, the results of the Einstein-Bumblebee spacetime cannot be directly derived as the limiting case of the Einstein-Bumblebee dS spacetime; therefore, in the Matrix Method adopted in this paper, we cannot obtain results for the $\Lambda = 0$ case. 
When the cosmological constant is small, the accuracy of the Matrix Method for $ N = 30 $ is poor, but this low accuracy can be improved by increasing the matrix order. 
However, when the cosmological constant is large, the Matrix Method exhibits very high accuracy, even with a low matrix order such as $ N = 15 $. 
Despite the potential for inherent errors in the numerical computation process and the limited number of parameters in the fitting procedure, the results obtained via the WKB Approximation and the Finite Difference Method meet the required accuracy.
Furthermore, our fitting results demonstrate consistency between the WKB Approximation and the Finite Difference Method.

\section{Conclusion}\label{sec5}

In this work, we investigated scalar field perturbations within the Einstein—Bumblebee theory by incorporating a cosmological constant. 
We derived the equation of motion for a massive scalar field, solved for the eigenfrequencies in the massive case using the matrix method, and verified the results in the massless case using the WKB approximation and finite difference method.
Moreover, we systematically analyzed the dependence of quasinormal mode frequencies on black hole parameters and numerically simulated dynamic waveforms characterizing scalar perturbation evolution in this spacetime.
Furthermore, the research methods and results of this paper could be extended to other black hole spacetimes within this theoretical framework, particularly to rotating black holes. 
Given the presence of superradiance phenomena in rotating black holes~\cite{Ding2020}, their dynamic stability remains a significant issue in black hole physics, which needs research and discussion in the future.

\section*{Data Availability}
The datasets used and/or analyzed during the current study are available from the corresponding author upon reasonable request.

\section*{Author Contributions}
Hao Hu conceived the research, performed data analysis, and wrote the original draft. 
Guoxiong Zhu supervised the research and assisted in the text revisions.
All authors contributed to the work.

\end{document}